\documentclass[aps,showpacs,pra,preprintnumbers,amsmath,amssymb,twocolumn]{revtex4-2}
\usepackage[utf8]{inputenc}
\usepackage[T1]{fontenc}
\usepackage{graphicx}
\usepackage{dcolumn}
\usepackage{appendix}
\usepackage{amsmath}
\usepackage{bm}
\usepackage[dvipsnames]{xcolor}
\usepackage{multirow}
\usepackage{pifont}
\usepackage{lipsum}

\newcommand{\me}{m_{\mathsf{e}}}

\usepackage{subfigure}
\usepackage{bbm}
\usepackage{enumerate}

\usepackage[
bookmarks=true,
colorlinks,
linkcolor=black,
urlcolor=black,
citecolor=black,
plainpages=false,
pdfpagelabels,
final,
breaklinks=true
]{hyperref}

\usepackage{titlesec}

\usepackage{array}

\usepackage{physics}

\begin{document}
	\allowdisplaybreaks
	
	\title{About the role of short and long trajectories on the quantum optical state after high-harmonic generation}
	
	\author{Javier Rivera-Dean}
	\affiliation{ICFO -- Institut de Ciencies Fotoniques, The Barcelona Institute of Science and Technology, 08860 Castelldefels (Barcelona)}
	\email{javier.rivera@icfo.eu}
	
	\date{\today}
	\begin{abstract}
		High-harmonic generation (HHG) involves the up-conversion of a high-intensity driving field into its harmonic orders.~This process is intrinsically non-classical, requiring from quantum mechanics for a complete explanation as, under suitable conditions, involves phenomena such as particle tunneling through a potential barrier.~When exposed to a high-intensity, low-frequency laser field, bound electrons ionize via tunneling, accelerate under the driving field, and recombine with the parent ion, emitting high-harmonic radiation.~However, electrons can follow two distinct pathways---short and long trajectories---during these steps.~In this work, we evaluate the signatures left by these trajectories on the quantum optical state after HHG, and observe that they lead to entanglement between the driving field and the generated harmonics.~By leveraging these correlations, we use harmonic generation to herald the creation of optical Schrödinger cat-like states in the driving field.~Additionally, using an ab-initio approach, we examine how propagation effects, which spatially separate the harmonic contributions from short and long trajectories, influence the non-classical characteristics of the emitted light.
	\end{abstract}
	\maketitle

\section{INTRODUCTION}

In physics, non-classical behaviors refer to phenomena that cannot be fully explained using classical theories, and therefore require a quantum mechanical framework for their understanding.~These behaviors underpin many physical processes at microscopic scales, yet they result in macroscopically distinguishable outcomes.~A prime example is the high-harmonic generation (HHG) process, where light from a highly intense, low-frequency input laser field is up-converted into radiation with frequencies that are multiples of the driving field~\cite{burnett_harmonic_1977,mcpherson_studies_1987,ferray_multiple-harmonic_1988}.~This up-conversion process can be controlled and manipulated to produce isolated~\cite{drescher_x-ray_2001} or sequences~\cite{antoine_attosecond_1996,paul_observation_2001} of attosecond-duration pulses, opening the door to the vast field of attosecond science~\cite{corkum_attosecond_2007,krausz_attosecond_2009,NobelPrize2023}.

From a microscopic perspective, HHG in atomic systems unfolds through a series of steps where an electron first ionizes, accelerates in the continuum under the influence of the input laser field, and finally recombines with the parent ion, emitting the accumulated energy in the form of high-harmonic radiation~\cite{krause_high-order_1992,corkum_plasma_1993,lewenstein_theory_1994}. Underlying these steps, non-classical behaviors emerge: to ionize and generate non-perturbative harmonic orders---beyond the scope of classical nonlinear optics---electrons must undergo tunneling events~\cite{chin_observation_1983}.~These tunneling events are possible thanks to the strong nature of the applied field, which bends the atomic potential the electron is subjected to, allowing it to tunnel and follow different pathways before recombination occurs~\cite{olga_simpleman,amini_symphony_2019}.

The HHG mechanism has been extensively studied from a semiclassical perspective, where the atomic system is treated quantum mechanically while the electromagnetic field is described classically.~However, the development of quantum optical descriptions has revealed non-classical features in the light's degrees of freedom~\cite{bhattacharya_stronglaserfield_2023,lewenstein_attosecond_2024,cruz-rodriguez_quantum_2024}.~Experimental~\cite{lewenstein_generation_2021,rivera-dean_strong_2022,lamprou_nonlinear_2023} and theoretical~\cite{stammer_high_2022,stammer_theory_2022,stammer_quantum_2023} studies have demonstrated that HHG can produce states of light in the driving field with seemingly non-classical characteristics, resembling those of optical Schrödinger cat states, when postselection operations are employed~\cite{rivera-dean_quantum_2024}.~Additionally, non-classical phenomena such as squeezing in the driving field have been shown to impact the output radiation~\cite{gorlach_high-harmonic_2023}.~In this direction, non-classical features have been theoretically reported when using bright squeezed vacuum states as main drivers~\cite{tzur_generation_2023}, and experimentally when using low intensity squeezed light as a probe~\cite{lemieux_photon_2024}.~Furthermore, theoretical studies driving HHG with more complex setups, have revealed the presence of single-mode and multi-mode squeezing. Specifically, when using driving fields of intensity sufficient to significantly deplete the atom~\cite{stammer_squeezing_2023}, in cavity QED setups resonant with the generated harmonic radiation~\cite{yi_generation_2024}, or in atoms initially driven to an excited state~\cite{rivera-dean_squeezed_2024}.

While these results demonstrate that HHG can serve as an important source of non-classical light, with frequencies spanning from the infrared to the extreme-ultraviolet regime~\cite{bhattacharya_stronglaserfield_2023,lewenstein_attosecond_2024,cruz-rodriguez_quantum_2024}, the influence of the non-classical electron dynamics inherent to the HHG process on the quantum optical features has not yet been fully explored.~Typically, the back-action of the electron dynamics in the continuum has been considered small and therefore neglected in many analyses~\cite{stammer_quantum_2023,gorlach_high-harmonic_2023}.~While other studies have accounted for this influence~\cite{stammer_squeezing_2023,rivera-dean_squeezed_2024}, these have been conducted within Markov-like approximations, which effectively neglect transient changes in the quantum optical state due to the electron's evolution during its excursion in the continuum.~However, it has been recently demonstrated that these quasi free-electron dynamics can lead to non-classical and non-Gaussian features in the light degrees of freedom~\cite{andrianov_formation_2024}.~In fact, in the context of above-threshold ionization (ATI) processes---where the electron ionizes driven by the strong field but misses recombination---it has been theoretically observed that these dynamics can lead to coherent state superpositions~\cite{rivera-dean_light-matter_2022}.

In this work, we theoretically investigate the impact of the non-classical electron dynamics during the HHG process on the output radiation.~We observe that, due to the electron's back-action while following distinct HHG pathways or quantum orbits---namely the short and long trajectories---the quantum optical state experiences shifts that depend on the specific trajectory taken by the electron.~This dynamic naturally introduces entanglement features between the field degrees of freedom, which we evaluate through the use of entanglement measures, and that become significant when multiple atoms contribute to the HHG process in an uncorrelated manner.~We demonstrate that, through heralding measurements, these quantum correlations can be harnessed to generate non-classical states of light in the form of optical Schrödinger cat-like states in the driving field which are, by construction, inherently different from those experimentally observed in Refs.~\cite{lewenstein_generation_2021,rivera-dean_strong_2022}.~Finally, using an ab-initio approach, we explore how propagation effects that spatially separate the harmonic contributions from short and long trajectories~\cite{bellini_temporal_1998,lynga_temporal_1999} influence the observed non-classical features.


\section{THEORY BACKGROUND}

To describe the influence of the electronic quantum orbits on the final quantum optical state, we begin by solving the Schrödinger equation governing the laser-matter interaction dynamics. In this section, we provide a concise overview, emphasizing on the physical significance of the obtained results and the approximations used.~A more detailed technical discussion, covering the mathematical nuances, is provided in Appendix~\ref{App:Solution}. For our initial state, we consider the joint electron-field system to be in the state $\ket{\Psi(t_0)} = \ket{\text{g}} \bigotimes_{q=1}^{q_c} \ket{\alpha \delta_{q,1}}$, where the electron is in its ground state, the driving field mode ($q=1$)---which we shall refer to as the fundamental mode---is in a coherent state of amplitude $\abs{\alpha}\gg 1$, and all harmonic modes ($q>1$) are in a vacuum state.

With this in mind, we express the time-dependent Schrödinger equation for the light-matter interaction, under the dipole approximation and within the single-active electron framework, as
\begin{equation}\label{Eq:TDSE}
	i\hbar \pdv{\lvert\bar{\Psi}(t)\rangle}{t}
		= \big[
				\hat{H}_{\text{at}} 
				+ \mathsf{e} 
					\big(
						E_{\text{cl}}(t) + \hat{E}(t)
					\big) \hat{r}
			\big] \lvert\bar{\Psi}(t)\rangle,
\end{equation}
where $\hat{H}_{\text{at}}$ is the atomic Hamiltonian, $\hat{E}(t) = -i \sum_{q=1}^{q_c} g(\omega_q)[\hat{a}_q e^{-i\omega_q t} - \hat{a}_q^\dagger e^{i\omega_q t}]$ the time-dependent electric field operator, with $\hat{a}^\dagger_q$ ($\hat{a}_q$) the creation (annihilation) operator acting on the $q$th harmonic mode, and $E_{\text{cl}}(t) = \Tr(\hat{E}(t) \bigotimes^{q_c}_{q=1} \dyad{\alpha \delta_{q,1}})$ the classical electric field.~Furthermore, we define $\ket{\Psi(t)} = e^{-i\hat{H}_{\text{field}}t/\hbar} \hat{D}_{q=1}(\alpha) \lvert \bar{\Psi}(t)\rangle$, with $\hat{H}_{\text{field}} = \sum_{q=1}^{q_c}\hbar \omega_q \hat{a}^\dagger_q\hat{a}_q$ the free-field Hamiltonian and $\hat{D}_q(\cdot)$ the displacement operator acting on the $q$th harmonic mode.~Under these definitions, the initial condition for Eq.~\eqref{Eq:TDSE} is given by $\lvert \bar{\Psi}(t_0)\rangle = \ket{\text{g}}\otimes \ket{\bar{0}}$, with $\ket{\bar{0}} \equiv \bigotimes_{q=1}^{q_c}\ket{0}$ representing the vacuum state for all modes.

It is important to note that we limit our analysis to a discrete set of modes located at harmonic frequencies of the fundamental frequency $\omega_L$, rather than considering a continuous spectrum.~This does not only simplify the analytical treatment but also implies that $g(\omega_q)$---a prefactor accounting for the laser-matter coupling arising from the expansion of the electric field operator into the creation and annihilation operators~\cite{ScullyBookCh1}---is a perturbative yet finite quantity that depends on the quantization volume: a smaller volume results in fewer available modes, while larger volumes yield a denser mode distribution.~However, despite the perturbative nature of $g(\omega_q)$---estimated to be in the order of $10^{-8}$ a.u.~for typical strong-field parameter regimes~\cite{rivera-dean_light-matter_2022}---when multiplied by the coherent state amplitude of the fundamental mode $\abs{\alpha}$, it results in electric field amplitudes on the order of $10^7$ V/m, which are sufficiently strong to drive HHG processes.~Moreover, this discrete-mode approximation is suitable for HHG, which typically involves a discrete emission spectrum.~The use of a continuous spectrum is not expected to significantly modify the results~\cite{stammer_quantum_2023}.

Given 
that Eq.~\eqref{Eq:TDSE} describes the evolution of an electron subjected to a strong-classical electromagnetic field $E_{\text{cl}}(t)$, while also accounting for the quantum optical fluctuations involved in the interaction through the coupling to $\hat{E}(t)$, we consider an extension of the standard Strong Field Approximation (SFA) as described in Refs.~\cite{lewenstein_theory_1994,amini_symphony_2019}. Our modified SFA involves using an ansatz of the form
\begin{equation}\label{Eq:ansatz:SFA}
	\lvert \bar{\Psi}(t)\rangle
		= a(t) \ket{\text{g}}\otimes \ket{\Phi_{\text{g}}(t)}
			+ \int \dd v \ b(v,t) \ket{v} \otimes \ket{\Phi(v,t)},
\end{equation}
where, following the semiclassical version of the SFA, we neglect the contribution of all bound states except for the electronic ground state.~Similarly, the continuum states are taken from the basis of scattering states $\{\ket{v}\}$, which satisfy $\hat{H}_{\text{at}}\ket{v} = m_{\mathsf{e}}v^2/2\ket{v}$, and lead to matrix elements of the position operator of the form $\mel{v}{\hat{r}}{v'} = i\hbar \pdv*{[\delta(v-v')]}{v} + \hbar f(v,v')$.~Here, $f(v,v')$ accounts for the part responsible for rescattering effects in ATI~\cite{amini_symphony_2019,stammer_quantum_2023}, which we do not consider in this analysis.

However, unlike in semiclassical analyses, our generalized ansatz in Eq.~\eqref{Eq:ansatz:SFA} explicitly includes the quantum optical components $\ket{\phi_{\text{g}}(t)}$ and $\ket{\Phi(v,t)}$ for the ground and continuum states, respectively. This provides a more comprehensive ansatz within the standard SFA framework as we assume, in general, $\ket{\Phi_{\text{g}}(t)} \neq \ket{\Phi(v,t)}$.~By setting $\ket{\Phi_{\text{g}}(t)} = \ket{\Phi(v,t)} = \ket{\bar{0}}$, we can recover the semiclassical results, while the formulation also accommodates the description of arbitrarily entangled light-matter states.~Thus, the two terms in Eq.~\eqref{Eq:ansatz:SFA} describe strong-field-driven processes where the electron ends up either in the ground or the continuum states, and the corresponding influence these dynamics have on the quantum optical state.~Therefore, it naturally accommodates both HHG and direct-ATI processes~\cite{rivera-dean_light-matter_2022}.

By inserting this ansatz in Eq.~\eqref{Eq:TDSE}, and projecting onto $\ket{\text{g}}$ and $\ket{\text{v}}$, we obtain the following set of coupled differential equations
\begin{align}
	 i \hbar \pdv{}{t}
		\big(
			a(t) \ket{\Phi_{\text{g}}(t)}
		\!\big)
		&= -I_p a(t)\ket{\Phi_{\text{g}}(t)}\nonumber
			\\&\quad+ \mathsf{e} \big( E_{\text{cl}}(t) + \hat{E}(t)\big) \label{Eq:Diff:ground}
		\\&\quad\quad \times
		\int \dd v \mel{\text{g}}{\hat{r}}{v} b(v,t) \ket{\Phi(v,t)},\nonumber
\end{align}

\begin{align}
		i\hbar \pdv{}{t}
		\big(
		b(v,t)\ket{\Phi(v,t)}
		\!\big)
		&= \dfrac{v^2}{2\me}b(v,t) \ket{\Phi(v,t)}\nonumber
		\\&\quad
		+ \mathsf{e}
		\big(E_{\text{cl}}(t) + \hat{E}(t)\big)\nonumber
		\\&\quad\quad\times 
		\mel{v}{\hat{r}}{\text{g}}
		a(t)
		\ket{\Phi_{\text{g}}(t)}\label{Eq:Diff:excited}
		\\&\quad
		+ i \hbar \mathsf{e}\big(E_{\text{cl}}(t) + \hat{E}(t)\big)\nonumber
		\\&\quad\quad\times
		\pdv{}{v}\big(b(v,t)\ket{\Phi(v,t)}\!\big),\nonumber
\end{align}
where $I_p$ denotes the ionization potential of the ground state. In Eq.~\eqref{Eq:Diff:ground}, the backaction of the electron dynamics on the quantum optical state is mediated solely by transitions from the ground to the continuum, represented by the second term on the right-hand side. In contrast, Eq.~\eqref{Eq:Diff:excited} reveals two possible pathways for this interaction. The first pathway, corresponding to the second term on the right hand side of Eq.~\eqref{Eq:Diff:excited}, is similar to that described for Eq.~\eqref{Eq:Diff:ground}, involving transitions between the ground and continuum states.~The second pathway, represented by the third term on the right-hand side of Eq.~\eqref{Eq:Diff:excited}, incorporates both the backaction of the electron on the quantum optical state (through the $\hat{E}(t)\pdv*{b(v,t)}{t}$ term, which resembles the interaction of a charge current with the electric field operator) and the influence of the quantum optical fluctuations on the electron dynamics (captured by the $\hat{E}(t)\pdv*{\ket{\Phi(v,t)}}{t}$ term). 

Regarding this last term, its contribution is expected to remain almost negligible, at least when considering coherent states of light as drivers of HHG processes. This expectation is based on the fact that semiclassical HHG analyses have been highly successful in describing the strong-field driven electron dynamics~\cite{shafir_resolving_2012,pedatzur_attosecond_2015,kneller_look_2022}. Moreover, recent results suggest that modifications to electron dynamics only become significant when the driving laser field exhibits non-classical features~\cite{even_tzur_photon-statistics_2023}. Based on this, we approximate Eq.~\eqref{Eq:Diff:excited} as follows
\begin{align}
		i\hbar \pdv{}{t}
		\big(
		b(v,t)\ket{\Phi(v,t)}
		\!\big)
		&\approx \dfrac{v^2}{2\me}b(v,t) \ket{\Phi(v,t)}\nonumber
		\\&\quad
		+ \mathsf{e}
		\big(E_{\text{cl}}(t) + \hat{E}(t)\big)\nonumber
		\\&\quad\quad\times 
		\mel{v}{\hat{r}}{\text{g}}
		a(t)
		\ket{\Phi_{\text{g}}(t)}\label{Eq:Diff:excited:approx}
		\\&\quad
		+ i \hbar \mathsf{e}\big(E_{\text{cl}}(t) + \hat{E}(t)\big)\nonumber
		\\&\quad\quad\times
		\pdv{b(v,t)}{v}\ket{\Phi(v,t)},\nonumber
\end{align}
which retains the back-action of the electron on the quantum optical state of the field, while constraining the electronic trajectories to those compatible with semiclassical dynamics.

By solving the set of differential equations in Eqs.~\eqref{Eq:Diff:ground} and \eqref{Eq:Diff:excited:approx} (see Appendices~\ref{App:Sol:Exc} and \ref{App:Sol:ground}), we can distinguish three distinct contributions to the final state of the joint electron-light system
\begin{equation}\label{Eq:Total:state}
	\lvert\bar{\Psi}(t)\rangle	
		= \lvert \bar{\Psi}_0(t) \rangle
			+ \lvert \bar{\Psi}_{\text{ATI}}(t) \rangle
			+ \lvert \bar{\Psi}_{\text{HHG}}(t) \rangle,
\end{equation} 
where the first contribution, $\lvert \bar{\Psi}_0(t)\rangle \approx e^{iI_p(t-t_0)/\hbar}\ket{\text{g}}\otimes \ket{\bar{0}}$, corresponds to events where the electron does not interact with the field and thus remains in the ground state.~The second contribution, $\lvert \bar{\Psi}_{\text{ATI}}(t)\rangle$, represents events where the electron ionizes directly to the continuum without returning to the parent ion, characterizing ATI events. The third contribution, $\lvert \bar{\Psi}_{\text{HHG}}(t)\rangle$, describes events where the electron returns to the parent ion, recombines with it, and emits high-harmonic orders of the fundamental mode.

It was demonstrated in Ref.~\cite{rivera-dean_light-matter_2022} that, due to the back-action of the electronic motion within the continuum on the field modes, the ATI contribution can result in light-matter entangled states, where the electronic ionization characteristics influence the final state of the field.~In this work, we focus on the HHG contribution, which can be approximated as (see Appendix~\ref{App:Final:state})
\begin{equation}\label{Eq:HHG:state}
	\lvert \bar{\Psi}_{\text{HHG}}(t)\rangle
		\approx \sum_{q=1}^{q_c} 
		\int \dd \boldsymbol{\theta}
			M_q(\boldsymbol{\theta})
				\hat{\vb{D}}
					\big(
						\boldsymbol{\delta}(\boldsymbol{\theta})
					\big)
					\hat{a}^\dagger_q
				\ket{\text{g}}\otimes \ket{\bar{0}},
\end{equation}
where we have used $\boldsymbol{\theta} \equiv (t_2,p,t_1)$ as a shorthand notation, with $p = m_{\mathsf{e}} v - \mathsf{e}A_{\text{cl}}(t)$ denoting the canonical momentum, $A_{\text{cl}}(t)$ the classical vector potential $E_{\text{cl}}(t) = - \pdv*{A_{\text{cl}}(t)}{t}$ and $\int \dd \boldsymbol{\theta} \equiv \int^t_{t_0}\dd t_2 \int \dd p \int^{t_2}_{t_0}\dd t_1$.~As a consequence of the HHG process, Eq.~\eqref{Eq:HHG:state} shows the generation of harmonic radiation across the entire spectral region.~Each harmonic contribution is weighted by two terms.~The first, $M_q(\boldsymbol{\theta})$ denotes the probability amplitude of finding an ionization, acceleration and recombination mechanism---happening at the conditions specified by the parameter $\boldsymbol{\theta}$---leading to the generation of the harmonic mode $q$. This quantity is given by
\begin{equation}\label{Eq:Prob:Amp}
	\begin{aligned}
	M_q(\boldsymbol{\theta})
		&= \dfrac{g(\omega_q)}{\hbar^2}e^{iI_p(t-t_0)/\hbar}
			e^{-iS^{(q)}_{\text{sc}}(\boldsymbol{\theta})/\hbar}
			\mel{\text{g}}{\hat{r}}{p+\mathsf{e}A_{\text{cl}}(t_2)}
			\\&\quad\times
			E_{\text{cl}}(t_1)
			\mel{p+\mathsf{e}A_{\text{cl}}(t_1)}{\hat{r}}{\text{g}},
	\end{aligned}
\end{equation}
where $S_{\text{sc}}^{(q)}(\boldsymbol{\theta})$ denotes the semiclassical HHG action~\cite{lewenstein_theory_1994,amini_symphony_2019}
\begin{equation}\label{Eq:SC:phase}
	S_{\text{sc}}^{(q)}(\boldsymbol{\theta})
		\equiv \dfrac{1}{2m_{\mathsf{e}}}
			\int^{t_2}_{t_1} \!\!\dd \tau \big[p+\mathsf{e}A_{\text{cl}}(\tau)\big]^2
			+ I_p(t_2-t_1) - \hbar q\omega_L t_2.
\end{equation}

The second contribution, $\hat{\boldsymbol{D}}(\boldsymbol{\delta}(\boldsymbol{\theta})) \equiv \prod^{q_c}_{q=1}\hat{D}_q(\delta_q(\boldsymbol{\theta}))$, represents a multimode displacement operator that shifts each harmonic mode by a quantity $\delta_q(\boldsymbol{\theta})$, which depends on the specific ionization conditions defined by the integration variable $\boldsymbol{\theta}$. The displacement is given by
\begin{equation}
	\delta_q(\boldsymbol{\theta})
		= \dfrac{\mathsf{e}}{\hbar} g(\omega_q)
			\int_{t_0}^t \dd \tau \Delta r(p,t,t_0) e^{iq\omega_L \tau},
\end{equation}
that is, as the Fourier transform of the spatial displacement performed by the electron between times $t_1$ and $t_2$, i.e., $\Delta r(p,t_2,t_1) = (1/m_{\mathsf{e}}) \int^{t_2}_{t_1} \dd \tau [p+\mathsf{e}A_{\text{cl}}(\tau)]$. Thus, Eq.~\eqref{Eq:HHG:state} provides a complete, within the considered approximations, description of the electronic and quantum optical dynamics occurring during the HHG process.~At time $t_1$, the electron transitions from the ground to the continuum state due to the light-matter interaction.~From $t_1$ to $t_2$, it propagates in the continuum, leading to an oscillating charge current due to its interaction with the classical electromagnetic field $E_{\text{cl}}(t)$, and thereby leads to a displacement in the field modes. Finally, the electron returns to the ground state at time $t_2$, emitting a harmonic photon in the process. Beyond this mechanism, the electron can undergo other processes, including multiple recombination steps within the duration of the applied field, each leading to higher-order contributions in $g(\omega_q)$, treated here as our perturbation parameter.~However, we find that the mechanism presented in Eq.~\eqref{Eq:HHG:state} is the dominant one (see Appendix~\ref{App:Final:state}).

\subsection{The saddle-point approximation}
The saddle-point approximation simplifies the evaluation of rapidly oscillating integrals by approximating them as a sum over carefully chosen points, obtained after suitably modifying the integration contour.~This technique is commonly employed in the analysis of strong-field processes, as high-order harmonics naturally produce fast-oscillating functions~\cite{lewenstein_theory_1994,amini_symphony_2019,nayak_saddle_2019}.~By closely examining Eq.~\eqref{Eq:HHG:state}, we find that this is indeed relevant to our analysis, as the probability amplitude in Eq.~\eqref{Eq:Prob:Amp} resembles the expression typically used to calculate the HHG spectrum, where the saddle-point approximation is often applied~\cite{lewenstein_theory_1994,olga_simpleman,nayak_saddle_2019,amini_symphony_2019}.

However, there is a crucial difference in our case compared to standard semiclassical analyses: in Eq.~\eqref{Eq:HHG:state}, we encounter the presence of a displacement operator that depends on the integration variables $\boldsymbol{\theta}$.~When expanded in a $\boldsymbol{\theta}$-independent basis set, such as the Fock basis, this contribution can introduce additional terms with rapidly oscillating behavior.~Nonetheless, since $\delta_q(\boldsymbol{\theta})$ is small---proportional to $g(\omega_q)$, our perturbation parameter---the dominant contributions from this expansion do not exhibit a rapidly oscillating behavior (see  Appendix~\ref{App:Saddle:Point}). Consequently, one can implement the saddle-point approximation as typically done in semiclassical analyses, by focusing solely on the rapidly oscillating behavior of $M(\boldsymbol{\theta})$.~This implies that the ``carefully chosen points'' refer to the saddle-points of Eq.~\eqref{Eq:SC:phase}, hereupon denoted as $\{\boldsymbol{\theta}_s\} \equiv \{(t_{\text{re}},p_s,t_{\text{ion}})\}$, which represent the complex-valued recombination time, canonical momentum and ionization time defining the HHG quantum orbits~\cite{lewenstein_theory_1994,amini_symphony_2019,nayak_saddle_2019}.~Consequently, Eq.~\eqref{Eq:HHG:state} can be approximated as
\begin{equation}\label{Eq:HHG:state:SPA}
	\lvert\bar{\Psi}_{\text{HHG}}(t)\rangle
		\approx \sum_{q}^{q_c}\sum_{\{\boldsymbol{\theta_s}\}}
			\Tilde{M}_q(\boldsymbol{\theta}_s) 
			\hat{\vb{D}}
				\big( \boldsymbol{\delta}(\boldsymbol{\theta}_s)\big)
				\hat{a}^\dagger_q
			\ket{\text{g}}\otimes \ket{\bar{0}},
\end{equation}
where $\Tilde{M}_q(\boldsymbol{\theta}_s)\equiv G(\boldsymbol{\theta}_s)M_q(\boldsymbol{\theta}_s)$, with $G(\boldsymbol{\theta}_s)$ denoting additional prefactors that weight each of the saddle-points~\cite{lewenstein_theory_1994,nayak_saddle_2019}.~In this work, we compute the saddle-points $\{\boldsymbol{\theta}_s\}$ numerically as described in Appendix~\ref{App:Saddle:Point}.

Before proceeding further, it is important to note that the saddle-point approximation can only be applied in the high-harmonic regime, i.e., when Eq.~\eqref{Eq:SC:phase} leads to rapidly-oscillating functions in Eq.~\eqref{Eq:Prob:Amp}.~For the field parameters considered in this work, we find that this condition generally holds for $q\gtrapprox15$.~Thus, we perform our analysis accordingly.~Additionally, similar approximations, such as the stationary phase method, have been similarly applied in quantum optical treatments of strong-field physics phenomena, as is the case of Ref.~\cite{yi_generation_2024}.

\subsection{Behavior of the displacements for short and long trajectories}
\begin{figure}
	\centering
	\includegraphics[width = 1 \columnwidth]{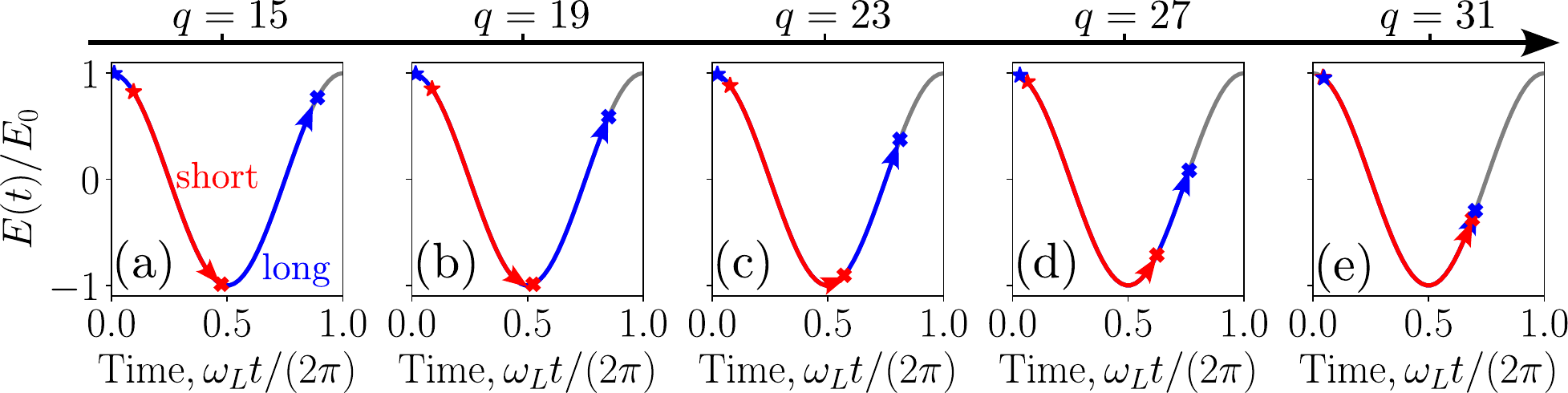}
	\caption{Pictorial representation of long (blue curve) and short (red curve) trajectories as a function of the harmonic modes. The trajectories are illustrated by the real parts of both the ionization (star markers) and recombination (cross markers) times.~These times were computed using $E_0 = 0.053$ a.u.~and $\omega_L =  0.057$ a.u., with the methods described in Appendix~\ref{App:Saddle:Point}.}
	\label{Fig:SC:traj}
\end{figure}
The HHG quantum orbits can typically be classified into two sets, referred to as short and long trajectories, depending on the time at which the electron recombines, as shown in Fig.~\ref{Fig:SC:traj}.~Here, we analyze how these trajectories affect the displacements $\{\delta_q(\boldsymbol{\theta}_s)\}$ that result from the electron's back-action on the quantum optical state.~In what follows, we denote the set of variables defining a short trajectory with the superscript $s$, i.e. $\boldsymbol{\theta}_s^{(s)}\equiv (t_{\text{re}}^{(s)},p^{(s)}_s,t_{\text{ion}}^{(s)})$, and those defining a long trajectory with the superscript $l$, i.e. $\boldsymbol{\theta}_s^{(l)}\equiv (t_{\text{re}}^{(l)},p^{(l)}_s,t_{\text{ion}}^{(l)})$.

\begin{figure}
	\centering
	\includegraphics[width=1\columnwidth]{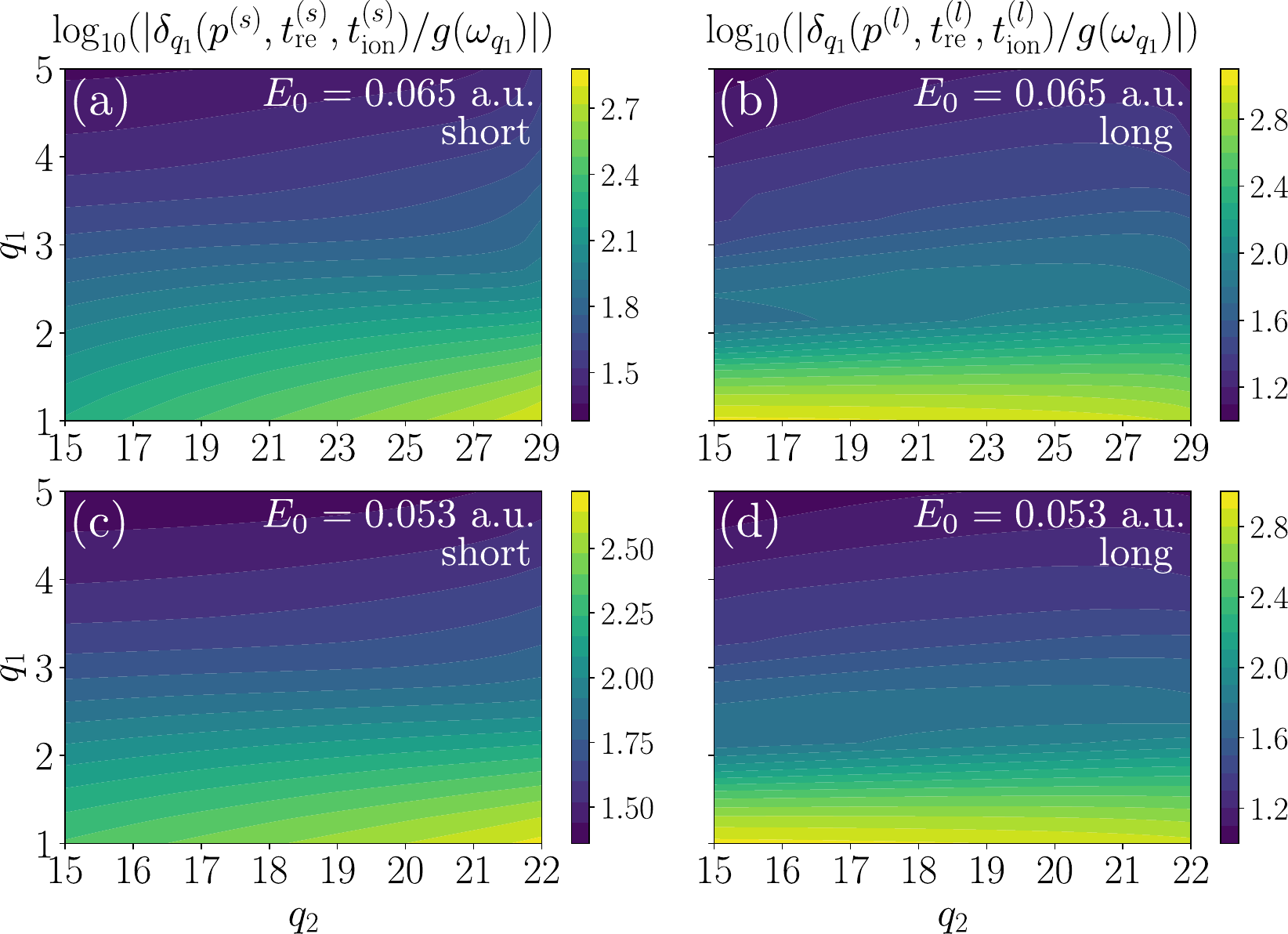}
	\caption{Evaluation of the absolute values of the displacement $\delta_{q_1}(\boldsymbol{\theta}_s)$ in units of $g(\omega_{q_1})$, displayed in logarithmic scale, for short (panels (a) and (c)) and long (panels (b) and (d)) trajectories.~The set of saddle-points $\{\boldsymbol{\theta}_s\}$ has been computed for different harmonic orders $q_2$ and evaluated for $\delta_{q_1}(\boldsymbol{\theta}_s)$ using various values of $q_1$, with $\omega_L = 0.057$ a.u.~in all cases.~In the upper panels, we set $E_0 = 0.065$ a.u.,~while to $E_0 = 0.053$ a.u.~in the lower panels.}
	\label{Fig:Disp:q1:q2}
\end{figure}

We begin by evaluating the strength of the displacement $\delta_{q}(\boldsymbol{\theta}_s)$ for the different harmonic modes.~To do so, we restrict ourselves to the case with a single pair of trajectories by considering a driving field of the form $E_{\text{cl}}(t) = E_0 \cos(\omega_L t)$.~We evaluate the saddle-points within the time span $t\in [0,2\pi/\omega_L]$. By fixing the values of $E_0$ and $\omega_L$, we compute the saddle-points~$\{\boldsymbol{\theta}_s^{(s)},\boldsymbol{\theta}_s^{(l)}\}$ for different harmonic modes $q_2$, and use them to compute $\delta_{q_1}(\boldsymbol{\theta}_s)$ for different values of $q_1$.

The results of this evaluation are presented in Fig.~\ref{Fig:Disp:q1:q2}, where we show $\log_{10}(\lvert\delta_{q_1}(\boldsymbol{\theta}_s)/g(\omega_L)\rvert)$ for various values of $q_1$ and $q_2$. We set $E_0 = 0.065$ a.u. in panels (a) and (b), which show the displacement obtained for the short and long trajectories respectively, and $E_0 = 0.053$ a.u. in panels (c) and (d), while fixing $\omega_L = 0.057$ a.u. in both cases.~In all scenarios, the most dominant contribution comes from the displacement of the fundamental mode, $q_1 = 1$. As $q_1$ increases, the resulting displacement decreases, being reduced by an order of magnitude already for $q_1 = 2$. Consequently, in the following, we focus only on the displacement obtained in the fundamental mode. We approximate $\hat{\vb{D}}(\boldsymbol{\delta}(\boldsymbol{\theta})) \approx \hat{D}_{q=1}(\delta_{q=1} (\boldsymbol{\theta}))\bigotimes_{q>1}\mathbbm{1} \equiv \hat{D}(\delta(\boldsymbol{\theta}))$, such that Eq.~\eqref{Eq:HHG:state:SPA} reads
\begin{equation}\label{Eq:HHG:SPA:fund}
	\lvert\bar{\Psi}_{\text{HHG}}(t)\rangle
		\approx \sum_{q}^{q_c}\sum_{\{\boldsymbol{\theta_s}\}}
			\Tilde{M}_q(\boldsymbol{\theta}_s) 
				\hat{D}
				\big( \delta(\boldsymbol{\theta}_s)\big)\hat{a}^\dagger_q
			\ket{\text{g}}\otimes \ket{\bar{0}}.
\end{equation}

\begin{figure}
	\centering
	\includegraphics[width=1\columnwidth]{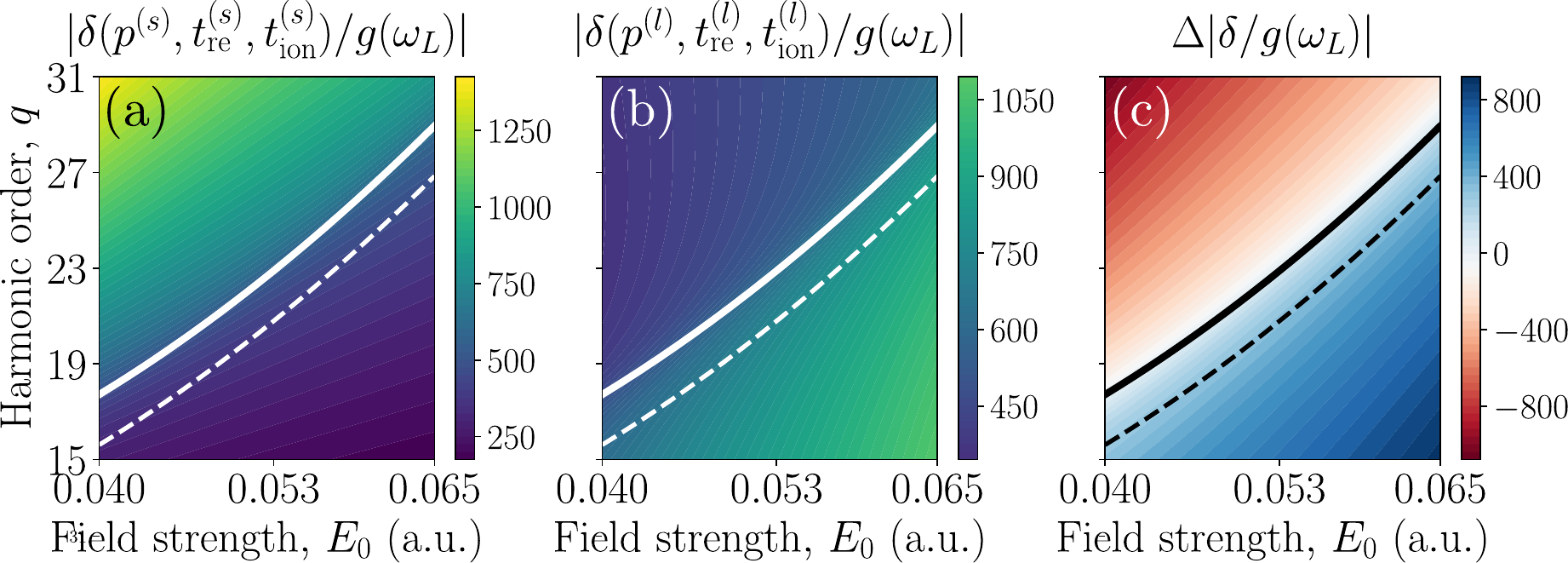}
	\caption{Panels (a) and (b) display the absolute value of the displacement $\delta(\boldsymbol{\theta}_s)$ for the fundamental mode, expressed in units of $g(\omega_L)$, for short and long trajectories, respectively.~Panel (c) shows the absolute value of the difference between the displacements obtained in both trajectories. The dashed curve indicates the harmonic cutoff for each field strength, while the solid curve denotes the position of the Stokes transition. The value $\omega_L = 0.057$ a.u. has been set in all cases.}
	\label{Fig:Disp:Field}
\end{figure}

Focusing only on the displacement affecting the harmonic mode $q=1$, Fig.~\ref{Fig:Disp:Field} shows how this quantity behaves as a function of the applied field strength, with $\omega_L = 0.057$ a.u.~fixed in all cases.~Specifically, panels (a) and (b) display the results for short and long trajectories, respectively, while panel (c) presents the difference between the two, defined as $\Delta\abs{\delta} = \lvert\delta(\boldsymbol{\theta}^{(l)}_s) -\delta(\boldsymbol{\theta}_s^{(s)})\rvert$. In all plots, the dashed curve represents the harmonic cutoff, while the solid curve represents the Stokes transition---the harmonic order for which $\text{Re}[S_{\text{sc}}(\boldsymbol{\theta}_s^{(s)})] = \text{Re}[S_{\text{sc}}(\boldsymbol{\theta}_s^{(l)})]$.

By comparing panels (a) and (b), we observe that below the Stokes transition, long trajectories provide a more significant displacement contribution than short trajectories.~This result is expected, as electrons associated with long trajectories spend more time propagating in the continuum, thus leading to a larger displacement contribution.~Furthermore, each of these contributions increases with the applied electric field.~As $q$ approaches the harmonic cutoff, the displacements for both trajectories converge, as the two trajectories begin to coalesce (see Fig.~\ref{Fig:SC:traj}).~Beyond the Stokes transition, the contribution from short trajectories diverges, leading to unphysical solutions~\cite{figueira_de_morisson_faria_high-order_2002,milosevic_role_2002,pisanty_imaginary_2020}. Therefore, beyond the Stokes transition, we neglect the short trajectory contributions and consider only the long trajectory contributions.

\subsection{A many atom extension}
One of the most interesting aspects about the results discussed so far is that the short and long trajectories naturally result in different displacements on the fundamental mode when considering harmonic orders within the plateau region, that is, below the cutoff.~Consequently, if we consider events leading to the generation of a single photon in the harmonic order $q\neq1$ with all other modes $q'\neq\{1,q\}$ in a vacuum state, and project Eq.~\eqref{Eq:HHG:SPA:fund} with respect to $\ket{1_q,\bar{0}_{q'}}$, we obtain
\begin{equation}\label{Eq:Schrödinger:cat}
	\begin{aligned}
	&\bra{1_q,\bar{0}_{q'}}\bar{\Psi}_{\text{HHG}}(t)\rangle
	\\&\hspace{0.5cm}
		= \Big[
				\Tilde{M}_q(\boldsymbol{\theta}^{(s)}_{s})
				\hat{D}\big(\delta(\boldsymbol{\theta}^{(s)}_{s})\big)
				+ \Tilde{M}_q(\boldsymbol{\theta}^{(l)}_{s})
				\hat{D}\big(\delta(\boldsymbol{\theta}^{(l)}_{s})\big)
			\Big]\!\ket{0},
	\end{aligned}
\end{equation}
that is, it results in a superposition of two different coherent states. However, unlike the approach in Ref.~\cite{lewenstein_generation_2021} where the non-classical states arise from the application of postselection techniques to the measured data~\cite{rivera-dean_quantum_2024}, in this case, they emerge due to the entanglement between the fundamental mode and the harmonics, established during the HHG process.

However, within the single-atom regime, these displacements are very small, leading to vanishing entanglement features~\cite{lewenstein_generation_2021,rivera-dean_strong_2022,stammer_quantum_2023} (see Appendix~\ref{App:Product:coh}).~To observe more significant values, one must instead consider the contribution of many atoms in the HHG process. To proceed, we first project the more general Eq.~\eqref{Eq:Total:state} onto those events where the electron ends up in the ground state of the system, leading to the following quantum optical state
\begin{equation}
	\lvert\bar{\Phi}_{\text{g}}(t)\rangle
		= \langle \text{g}\vert \bar{\Psi}(t)\rangle
		= \langle \text{g}\vert  \bar{\Psi}_0(t)\rangle
			+ \langle \text{g}\vert  \bar{\Psi}_{\text{HHG}}(t)\rangle,
\end{equation}
which in terms of Eq.~\eqref{Eq:HHG:SPA:fund} and up to a phase prefactor, can be written as
\begin{equation}\label{Eq:ground:state:cond}
	\lvert\bar{\Phi}_{\text{g}}(t)\rangle
		\approx
			\Bigg[
				\mathbbm{1}
				+ \sum_{q}^{q_c}\sum_{\{\boldsymbol{\theta_s}\}}
					\bar{M}_q(\boldsymbol{\theta}_s) 
					\hat{\vb{D}}
					\big( \boldsymbol{\delta}(\boldsymbol{\theta}_s)\big)\hat{a}^\dagger_q
			\Bigg] \ket{\bar{0}},
\end{equation}
where $\bar{M}_q(\boldsymbol{\theta}) = e^{-iI_p(t-t_0)/\hbar} \Tilde{M}_q(\boldsymbol{\theta})$.~Then, assuming that the time-dependent dipole moments of the different atoms involved in the HHG process are uncorrelated~\cite{sundaram_high-order_1990}, we can approximate the quantum optical state in the many-atom regime as follows~\cite{rivera-dean_squeezed_2024}
\begin{equation}\label{Eq:Many:Atom:State}
		\lvert\bar{\Phi}_{\text{g}}(t)\rangle
			\approx
				\Bigg[
					\mathbbm{1}
					+ \sum_{q}^{q_c}\sum_{\{\boldsymbol{\theta_s}\}}
					\bar{M}_q(\boldsymbol{\theta}_s) 
					\hat{D}
						\big( 		\delta(\boldsymbol{\theta}_s)\big)\hat{a}^\dagger_q
				\Bigg]^{N_{\text{at}}} \ket{\bar{0}},
\end{equation}
where $N_{\text{at}}$ represents the total number of atoms participating in the process.

\section{RESULTS}
In this section, we discuss non-classical features that arise from Eq.~\eqref{Eq:Many:Atom:State}. However, before proceeding, we need to make few considerations. In typical HHG setups, the number of atoms in the harmonic generation region can easily exceed $N_{\text{at}} = 10^{12}$. This makes the numerical analysis of Eq.~\eqref{Eq:Many:Atom:State} quite challenging, as it involves applying an operator that entangles different field degrees of freedom $N_{\text{at}}$ times.~To make the analysis feasible, we reduce the number of atoms to the $N_{\text{at}}\propto 10^5$ regime while proportionally increasing the light-matter coupling factor $g(\omega_L)$ to $g(\omega_L) \propto 10^{-3}$~\cite{andrianov_formation_2024}.~This compensates for the reduced number of atoms while ensuring that the overall interaction strength is maintained. We use hydrogen atom parameters in our analysis and set $I_p = 0.5$ a.u., consequently.

To reduce the effective number of Hilbert spaces we need to handle numerically, we restrict our analysis to only two modes: the fundamental and a harmonic mode $q$ with $q\geq 15$, ensuring the saddle-point approximation is applicable.~This effectively assumes that all modes $q'\neq\{1,q\}$ remain in the vacuum state.~By projecting Eq.~\eqref{Eq:Many:Atom:State} onto this subspace, we obtain
\begin{equation}\label{Eq:two:harm:state}
	\lvert\bar{\Phi}^{(q)}_{\text{g}}(t)\rangle
		\approx
			\Bigg[
			\mathbbm{1}
			+ \sum_{\{\boldsymbol{\theta_s}\}}
				\bar{M}_q(\boldsymbol{\theta}_s) 
				\hat{D}
				\big( \delta(\boldsymbol{\theta}_s)\big)\hat{a}^\dagger_q
			\Bigg]^{N_{\text{at}}} \ket{\bar{0}},
\end{equation}
with $	\lvert\bar{\Phi}^{(q)}_{\text{g}}(t)\rangle = \bra{\bar{0}_{q'\neq\{1,q\}}}\bar{\Phi}_{\text{g}}(t)\rangle$.~More details about the numerical evaluation of the quantum optical observables and entanglement measures is provided in Appendix~\ref{App:Num:Eval}.

\subsection{Entanglement features between the different field modes}

In the previous section, we saw the the state of the harmonics and the fundamental are correlated, as we observed that the generation of different harmonic modes is associated with different displacements on the fundamental mode which, as  we have seen, vary for the short and long trajectories followed during the HHG process. If these displacements were vanishing, we would end up with a product state between fundamental and harmonics in Eq.~\eqref{Eq:two:harm:state} (see Appendix~\ref{App:Product:coh}), demonstrating a lack of quantum correlations between the fundamental mode and the harmonics.

\begin{figure}
	\centering
	\includegraphics[width=1\columnwidth]{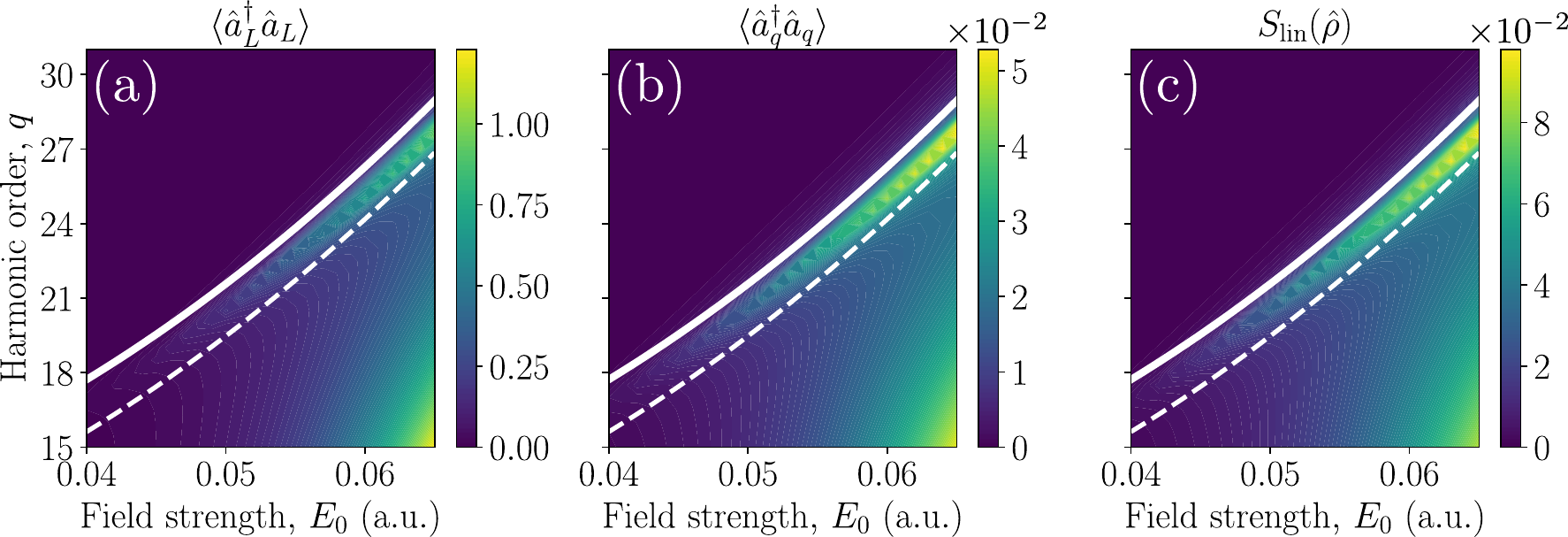}
	\caption{Mean photon number and linear entropy as a function of the field strength.~Panel (a) shows the mean photon number of the fundamental mode as a function of the harmonic mode that is generated, represented in the $y$-axis. Panel (b) displays the mean photon number of the various harmonic modes, denoted in the $y$-axis.~Panel (c) illustrates the linear entropy between the fundamental mode and he $q$th harmonic mode. In these calculations, $\omega_L = 0.057$ a.u., $g(\omega_L) = 5\times 10^{-3}$ and $N_{\text{at}} = 10^5$.}
	\label{Fig:Mel:entropy}
\end{figure}

However, when considering the contribution of many atoms, as in Eq.~\eqref{Eq:two:harm:state:traj} within the parameter regime outlined at the beginning of this section, we observe significant correlations between the fundamental mode and the harmonics.~Panels (a) and (b) in Fig.~\ref{Fig:Mel:entropy} illustrate these correlations by displaying the mean photon number of Eq.~\eqref{Eq:two:harm:state} for the fundamental mode and the harmonics, respectively, as a function of the field strength and for different harmonic orders. Here, we use $\omega_L = 0.057$ a.u., $N = 10^{5}$ and $g(\omega_L) = 5\times 10^{-3}$.~It is important to note that for the mean photon number of the fundamental mode shown in panel (a), this does not represent the total mean photon number, as we are working in a displaced frame with respect to the initial coherent state amplitude $\alpha$.~However, for the purposes of highlighting the existing correlations between the fundamental and harmonic modes, this is sufficient.

By comparing panels (a) and (b), we observe that the overall behavior is the same for both the mean photon number of the fundamental mode and the harmonics. Although they differ in magnitude, both quantities show maxima in the same regions.~Specifically, for high values of the field strength ($E_0 \gtrapprox 0.05$ a.u.), we observe the presence of maxima at $q < 18$ and close to the cutoff region.~The latter is motivated by the dominant probability of finding long trajectories in this region (see Appendix~\ref{App:Saddle:Point}). It is worth noting that this behavior---the similarity between panels (a) and (b)---aligns with previous experimental~\cite{tsatrafyllis_high-order_2017} and theoretical~\cite{gonoskov_quantum_2016,moiseyev_non-hermitian_2024} investigations where the harmonic radiation spectrum was inferred from the photon statistics of the IR driving laser field.~However, in our case, the presence of well-defined odd harmonic orders is not found as we are studying the influence of two trajectories obtained within one cycle of the field.~The use of multi-cycle laser fields should, however, give rise to this characteristic behavior of HHG spectra (see Appendix~\ref{App:Final:state}).

The behavior we have just described points toward the presence of quantum correlations between the different field modes, since the state of the fundamental mode is determined by that of the harmonics.~We quantitatively characterize the presence of such correlations in panel (c), by evaluating the linear entropy $S_{\text{lin}}(\hat{\rho}) = 1 - \tr(\hat{\rho}^2)$~\cite{agarwal_quantitative_2005,berrada_beam_2013}, a valid measure of measure when dealing with pure states.~This entanglement measure takes as input the partial trace of the state of interest with respect to one of the subsystems---either $\hat{\rho} = \tr_{q=1}(\lvert \bar{\Phi}_{\text{g}}^{(q)}(t)\rangle\!\langle\bar{\Phi}_{\text{g}}^{(q)}(t)\rvert)$ or $\hat{\rho} = \tr_{q\neq1}(\lvert \bar{\Phi}_{\text{g}}^{(q)}(t)\rangle\!\langle\bar{\Phi}_{\text{g}}^{(q)}(t)\rvert)$---and provides information about the amount of entanglement present, with $S_{\text{lin}}(\hat{\rho}) = 0$ indicating the absence of entanglement. 

As seen in panel (c), we indeed observe the presence of entanglement features which, although small, are significant---on the order of $10^{-2}$.~Moreover, when compared with panels (a) and (b), the overall behavior of this entanglement measure with respect to the field strength and the harmonic modes closely mirrors that of the mean photon number.~This similarity could suggest that correlated measurements between the intensity of specific harmonics and the corresponding shifts they induce in the fundamental mode could serve as an indicator of entanglement.~This is particularly interesting from the experimental point of view, as computing entanglement measures such as the linear entropy requires performing quantum tomography of the optical state, whereas the mean photon number can be measured using photodetectors sensitive to the spectral region of interest.

\begin{figure}
	\centering
	\includegraphics[width=1\columnwidth]{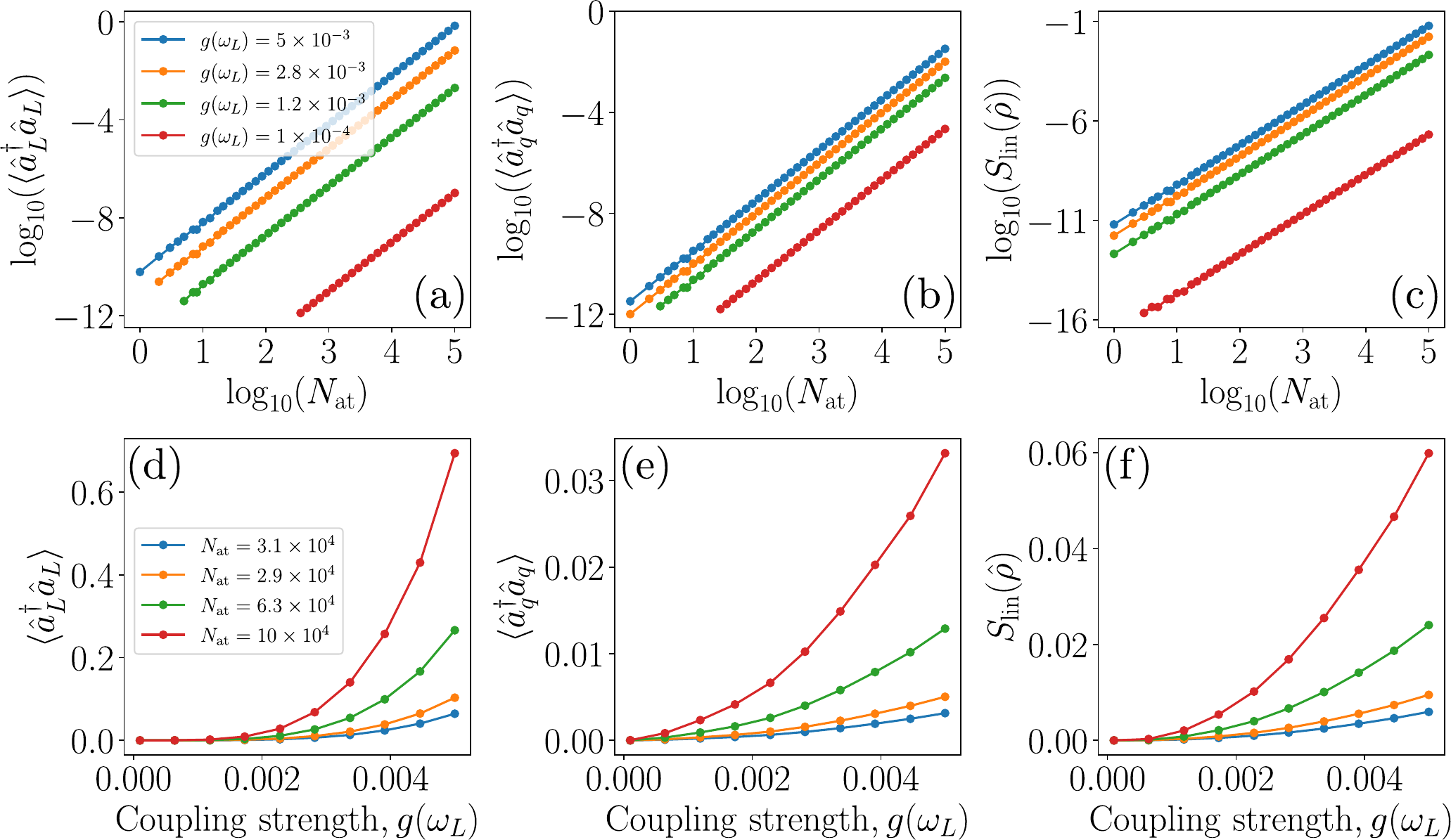}
	\caption{Mean photon number and linear entropy as a function of the coupling strength $g(\omega_L)$ and the number of atoms $N_{\text{at}}$. In the upper row, $g(\omega_L)$ is kept fixed, and the quantities are plotted as a function of $N_{\text{at}}$. In the lower row, $N_{\text{at}}$ is fixed, and the quantities are plotted as a function of $g(\omega_L)$. The results are shown for the 19th harmonic order, with $E_0 = 0.065$ a.u. and $\omega_L = 0.057$ a.u., though similar behaviors are observed across different parameter sets.}
	\label{Fig:Mel:entropy:diff:g:N}
\end{figure}

While the results in Fig.~\ref{Fig:Mel:entropy} have been obtained for fixed values of $N_{\text{at}}$ and $g(\omega_L)$, we expect these two quantities to significantly influence the correlations between the field modes. This is indeed demonstrated in Fig.~\ref{Fig:Mel:entropy:diff:g:N}, where we fixed $E_0 = 0.065$ a.u., $\omega_L = 0.057$ a.u., and $q=19$, although different parameters yield similar results.~Specifically, in the upper row we vary $N_{\text{at}}$ while keeping $g(\omega_L)$ fixed and examine how this influences the quantities mentioned earlier.~In the lower row, we vary $g(\omega_L)$ while keeping $N_{\text{at}}$ fixed. In both scenarios, we observe an exponential increase in all three evaluated quantities: the mean photon numbers for the fundamental and harmonic modes, and the linear entropy, from left to right.~This underscores the importance of these two parameters in determining whether significant levels of entanglement can be observed in the final state.

\subsection{Non-classical states of light heralded by harmonic generation}

The presence of entanglement between the fundamental and the harmonic modes opens up the possibility of designing more complex non-classical states of light by means of heralding measurements~\cite{ourjoumtsev_generating_2006,rivera-dean_squeezed_2024}.~By heralding measurements, we refer to conditional measurements done in one of the subsystems, in this case either on the fundamental on the harmonic modes, that depending on their outcome leave the other subsystem in a defined state.~Then, depending on the characteristics of the interaction, the resultant state can show non-classical features such as, for instance, the presence of Wigner negativities~\cite{hudson_when_1974}.

In our case, we observed in Eq.~\eqref{Eq:Schrödinger:cat} that detection of harmonic radiation could result in the generation of optical Schrödinger cat-like states.~Within the context of the two-mode formalism we have adopted in this section, this measurement can be formally written within the Positive Operator Valued Measured (POVM) formalism~\cite{NielsenBookCh1} as $\{\mathbbm{1}\otimes \dyad{0},\mathbbm{1}\otimes (\mathbbm{1} - \dyad{0})\}$, with the first mode representing the fundamental mode, while the second denoting the harmonic mode $q$ of interest. In this set, the first term denotes those cases where no harmonic radiation is detected, while the second one represents those cases where it indeed has been measured. Each of these outcomes will have associated a given probability. However, here we focus on those cases where the second outcome has been obtained---when radiation has been actually detected---and evaluate the quantum state of the fundamental mode afterwards, that is
\begin{equation}\label{Eq:Heralded:state}
	\hat{\rho}
		= \tr_q[
			\big(
				\mathbbm{1}
				\otimes
				(\mathbbm{1} - \dyad{0})
			\big) \lvert \bar{\Phi}_{\text{g}}^{(q)}(t)\rangle\!\langle \bar{\Phi}_{\text{g}}^{(q)}(t)\rvert
			].
\end{equation}

\begin{figure}
	\centering
	\includegraphics[width=1\columnwidth]{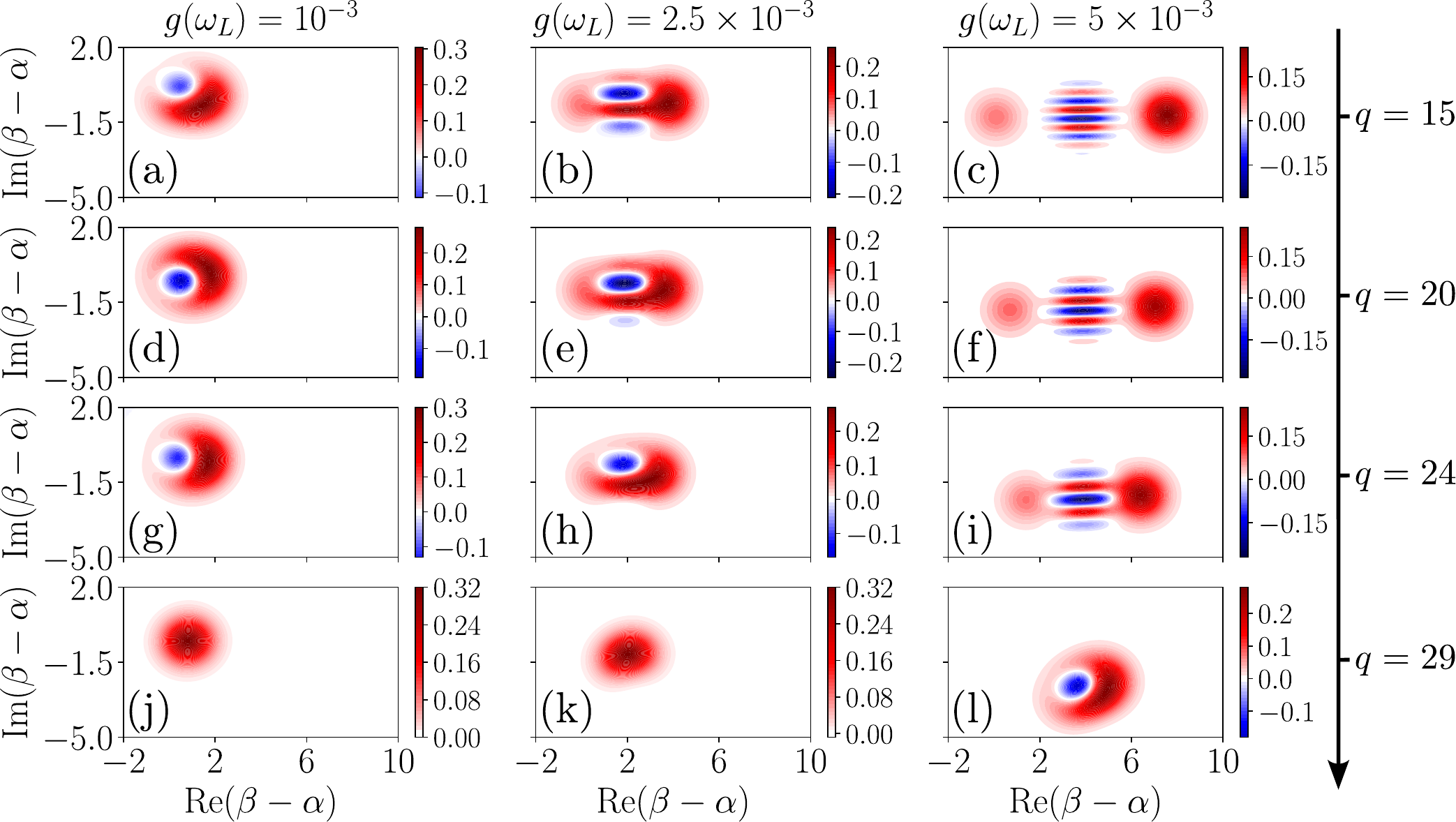}
	\caption{Wigner functions of the fundamental mode after heralding the generation of radiation in the $q$th harmonic mode, displayed from lower to higher harmonic orders across the rows. These calculations are performed with $E_0 = 0.065$ a.u. and $\omega_L = 0.057$ a.u., values at which the mean photon number and entanglement characteristics in Fig.~\ref{Fig:Mel:entropy} are most prominent.}
	\label{Fig:Wigners:herald}
\end{figure}

In Fig.~\ref{Fig:Wigners:herald}, we compute the Wigner function of Eq.~\eqref{Eq:Heralded:state} as $W(\beta) = \text{tr}( \hat{D}(\beta)\hat{\Pi}\hat{D}(-\beta)\hat{\rho})$~\cite{royer_wigner_1977} for different harmonic modes $q$ and coupling strengths $g(\omega_L)$. In all cases, we fix $E_0 = 0.065$ a.u.~and $\omega_L = 0.057$ a.u., as these values correspond to the highest displacement and entanglement features (see Fig.~\ref{Fig:Mel:entropy}).~Overall, we observe that the resulting states exhibit optical Schrödinger cat-like features in which one of the terms in the superpositions is more prominent than the other one.~The optical Schrödinger cat characteristics become more evident as the coupling strength increases, leading in the case of $g(\omega_L) = 5 \times 10^{-3}$ to two well-distinguished maxima at the edges of the distribution---the left one corresponding to short trajectories and the right peak to long trajectories---along with an interference pattern in between, signalling the presence of a quantum superposition.~Conversely, the optical cat state features diminish as $q$ increases.~This reduction is expected, as for harmonic orders approaching the cutoff (around $q\approx 27$), the two trajectories tend to coalesce, leading to similar displacements in the fundamental mode. 

Finally, it is important to highlight that, although similar in shape, the optical Schrödinger cat states obtained here are fundamentally different from those experimentally achieved in Refs.~\cite{lewenstein_generation_2021,rivera-dean_strong_2022} within the context of HHG. The states derived here are observed after the performance of heralding measurements on the harmonic modes, which physically implies that they will inevitably suffer from decoherence effects when propagating through any medium.~In contrast, the states in the aforementioned references are obtained through postselection on the measurement data, meaning that no non-classical state actually propagates through the experimental setup.~Instead, the non-classical statistics are derived from a postselection process after measurement~\cite{rivera-dean_quantum_2024}.~In this case, the resulting statistics are not affected by decoherence factors such as the strong neutral filters used in the measurements.~However, despite this advantage, it has been shown that non-classical features obtained via postselection might not be suitable for certain quantum information tasks, such as demonstrating Bell nonlocality~\cite{pearle_hidden-variable_1970} and related applications~\cite{brunner_bell_2014}.

\subsection{About the role of propagation effects}
Propagation effects of the generated radiation play a fundamental role in HHG. For instance, when atoms are excited with lasers that have an intensity-dependent spatial configuration---where atoms at different spatial points experience different intensities---the coherent summation of their contributions to the HHG radiation can yield numerically resolved harmonic spectra ~\cite{kim_enhanced_2000}. Without this coherent summation, the spectra would otherwise exhibit a quasi-continuous structure~\cite{salieres_temporal_1998,schafer_high_1997}. More relevant to the analysis presented in this work, propagation effects can cause spatial divergence between harmonics generated through short and long trajectories~\cite{bellini_temporal_1998,lynga_temporal_1999}. This makes possible to experimentally resolve their different contributions to the harmonic radiation by placing photodetectors at different spatial directions.

To understand the role that propagation effects could have on the non-classical features analyzed so far, we consider an ab-initio approach for their modeling. Specifically, we associate the harmonic modes generated through the different electron paths to distinct spatial modes. We label these modes as $\hat{a}_{q,s}$ ($\hat{a}^\dagger_{q,s}$) for the short trajectories and $\hat{a}_{q,l}$ ($\hat{a}^\dagger_{q,l}$) for the long trajectories. Consequently, we make the substitutions $\hat{D}(\delta(\boldsymbol{\theta}^{(s)}_s)) \hat{a}^\dagger_{q} \to \hat{D}(\delta(\boldsymbol{\theta}^{(s)}_s)) \hat{a}^\dagger_{q,s}$ and $\hat{D}(\delta(\boldsymbol{\theta}^{(l)}_s)) \hat{a}^\dagger_{q} \to \hat{D}(\delta(\boldsymbol{\theta}^{(l)}_s)) \hat{a}^\dagger_{q,l}$ to Eq.~\eqref{Eq:two:harm:state}, so that that it now reads
\begin{align}
	\lvert\bar{\Phi}^{(q)}_{\text{g}}(t)\rangle
	\approx
	\Bigg[
	\mathbbm{1}
	&+ 
	\bar{M}_q(\boldsymbol{\theta}^{(s)}_s) 
	\hat{D}
		\big(
			\delta(\boldsymbol{\theta}^{(s)}_s)
		\big)\hat{a}^\dagger_{q,s}\label{Eq:two:harm:state:traj}
	\\&+ 
	\bar{M}_q(\boldsymbol{\theta}^{(l)}_s) 
	\hat{D}
		\big(
			\delta(\boldsymbol{\theta}^{(l)}_s)
		\big)\hat{a}^\dagger_{q,l}
	\Bigg]^{N_{\text{at}}} \ket{\bar{0}}. \nonumber
\end{align}

\begin{figure}
	\centering
	\includegraphics[width=1\columnwidth]{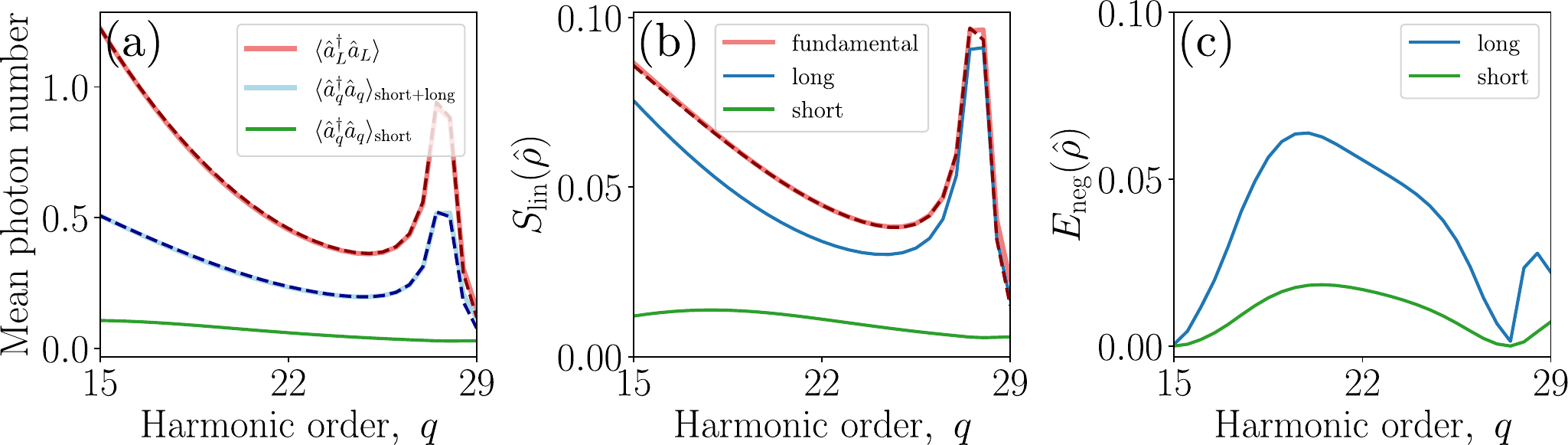}
	\caption{Mean photon number and entanglement characteristics using the ab-initio propagation model in Eq.~\eqref{Eq:two:harm:state:traj}. Panels (a) and (b) show the mean photon number and linear entropy respectively, with the dashed curves representing the results obtained in Fig.~\ref{Fig:Mel:entropy}. Panel (c) illustrates the entanglement between the fundamental and the harmonic modes, computed via the logarithmic negativity, for cases where heralding is performed on harmonics generated through the long (blue curve) or short (green curve) trajectories.~In these calculations, $E_0 = 0.065$ a.u., $\omega_L = 0.057$ a.u., $g(\omega_L) = 5\times10^{-3}$ and $N_{\text{at}} = 10^5$.}
	\label{Fig:Quantities:prop}
\end{figure}

In panels (a) and (b) of Fig.~\ref{Fig:Quantities:prop}, we evaluate how the spatial distinction between short and long trajectories affects the mean photon number and entanglement properties evaluated earlier.~We consider $E_0 = 0.065$ a.u. for the field strength, and $\omega_L = 0.057$ a.u. for the frequency. The solid curves represent results obtained using Eq.~\eqref{Eq:two:harm:state:traj}, while the dashed curves correspond to those using Eq.~\eqref{Eq:two:harm:state}. In panel~(a), where we show the mean photon number of the different harmonic modes and of the fundamental as a function of the harmonic mode that is generated, the blue solid curve represents the joint contribution of harmonics generated through both long and short trajectories, i.e., $\langle \hat{a}_q^\dagger \hat{a}_q \rangle =  \langle \hat{a}_{q,s}^\dagger \hat{a}_{q,s} \rangle + \langle \hat{a}_{q,l}^\dagger \hat{a}_{q,l} \rangle$. The green curve displays the contribution arising solely from the short trajectories.~As observed, both the red and blue curves exhibit similar features to those in Fig.~\ref{Fig:Mel:entropy}~(a), with the green curve indicating that the dominant contribution to the mean photon number arises from the long trajectories.

Similar features are obtained for the linear entropy in panel (b).~The red curve shows the entanglement between the fundamental and harmonics generated through both long and short trajectories.~The blue and green curves show the degree of entanglement in the two remaining bipartite configurations:~the entanglement between the joint system of the fundamental mode and harmonics from short trajectories with those from long trajectories, and the converse scenario, respectively.~Thus, when considering harmonics generated by both trajectories together, the entanglement with the fundamental mode remains consistent with the case without propagation (red dashed curve).~However, the physical distinction between the harmonic contribution of both trajectories due to propagation, allows us to characterize the entanglement separately.~This leads us to conclude that the strongest entanglement contribution arises from the long trajectory harmonics.

Although the overall entanglement features between the fundamental and harmonics do not significantly change, as shown in Fig.~\ref{Fig:Quantities:prop} panel (b), the physical distinction between the harmonics generated by short and long trajectories enables us to design different scenarios that can alter the entanglement characteristics with the harmonic modes.~For instance, in Fig.~\ref{Fig:Quantities:prop}~(c) we consider scenarios where we herald on the generation of the $q$th harmonic order from one of the trajectories, taking advantage of their spatial separation.~Using a similar POVM set as in the previous subsection, and given that now we are dealing with a tripartite scenario, the POVM set for heralding on harmonics generated via the short trajectories is given by $\{\mathbbm{1}_{q=1}\otimes \dyad{0}_s \otimes \mathbbm{1}_l, \mathbbm{1}_{q=1}\otimes (\mathbbm{1}_s - \dyad{0}_s) \otimes \mathbbm{1}_l \}$. Similarly, for the heralding on the long trajectories, the POVM set is $\{\mathbbm{1}_{q=1}\otimes\mathbbm{1}_s \otimes \dyad{0}_l, \mathbbm{1}_{q=1}\otimes \mathbbm{1}_s \otimes (\mathbbm{1}_l - \dyad{0}_l)\}$. Then, focusing on the outputs where harmonic radiation has been indeed detected, we obtain
\begin{equation}\label{Eq:herald:short}
	\hat{\rho}_l(t)
	= \tr_s[
	\big(
	\mathbbm{1}_{q=1}\otimes (\mathbbm{1}_s - \dyad{0}_s) \otimes \mathbbm{1}_l
	\big)
	\lvert \bar{\Phi}_{\text{g}}^{(q)}(t)\rangle\!\langle \bar{\Phi}_{\text{g}}^{(q)}(t)\rvert],
\end{equation}
for the joint state of the fundamental mode and harmonics obtained through long trajectories, while 
\begin{equation}\label{Eq:herald:long}
	\hat{\rho}_s(t)
		 = \tr_l[
		 		\big(
		 			\mathbbm{1}_{q=1} \otimes \mathbbm{1}_s \otimes (\mathbbm{1}_l - \dyad{0}_l)
		 		\big)
		 		\lvert \bar{\Phi}_{\text{g}}^{(q)}(t)\rangle\!\langle \bar{\Phi}_{\text{g}}^{(q)}(t)\rvert],
\end{equation}
for the joint state of the fundamental mode and harmonics obtained through short trajectories.

In panel (c) we show the amount of entanglement between the fundamental mode and the different harmonics for both Eqs.~\eqref{Eq:herald:short} (blue curve) and \eqref{Eq:herald:long} (green curve).~In this case, we use the logarithmic negativity $E_{\text{neg}}(\hat{\rho}) = \log_2(2N + 1)$ as the entanglement measure~\cite{peres_separability_1996,horodecki_separability_1996}, where $N$ represents the sum of all negative eigenvalues of the partially transposed density matrix with respect to one of the subsystems.~This measure is necessary in this case because the linear entropy is no longer a valid indicator of entanglement when dealing with mixed states, which generally arise after heralding operations.~Using this approach, we observe significant differences compared to the results in panel (b).~Specifically, the amount of entanglement is greatly suppressed for the lowest harmonic orders considered, as well as for harmonic orders near the cutoff.~Yet, in between these two ranges, entanglement is significantly enhanced compared to the case without heralding.~Furthermore, similar to panel (b), the contribution from the long trajectories dominates over that of the short trajectories.

\begin{figure}
	\centering
	\includegraphics[width=1\columnwidth]{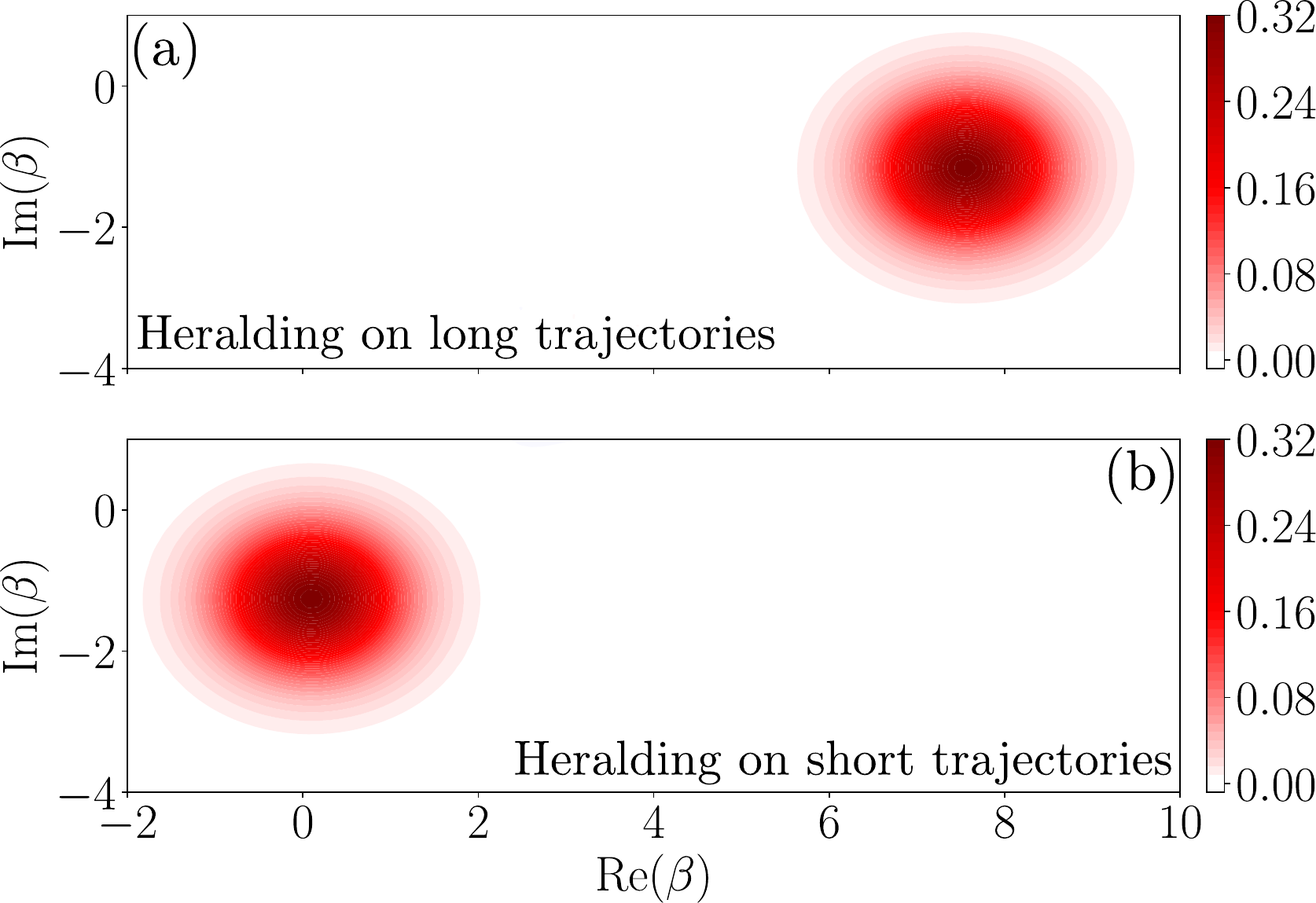}
	\caption{Wigner function of the fundamental mode after heralding the generation of radiation in the 20th harmonic mode, produced through long (panel (a)) or short (panel (b)) trajectories, using the ab-initio propagation model in Eq.~\eqref{Eq:two:harm:state:traj}. Calculations were performed with $E_0 = 0.065$ a.u., $\omega_L = 0.057$ and $g(\omega_L) = 5 \times 10^{-3}$.}
	\label{Fig:Wigners:prop}
\end{figure}

Finally, in Fig.~\ref{Fig:Wigners:prop} we evaluate how heralding operations affect the generation of non-classical states of light in the fundamental mode under the considered ab-initio propagation model.~The Wigner functions are evaluated for $q=20$, although similar results are obtained for other harmonic orders, using $g(\omega_L) = 5\times 10^{-3}$.~The results are shown for heralding on long (panel (a)) and short (panel (b)) trajectories.~In both cases, we observe an absence of non-classical features, with the final state exhibiting a Gaussian distribution in phase space without any negative regions.~This difference with respect to the results in Fig.~\ref{Fig:Wigners:herald} can be understood from the fact that propagation leads to a physical distinction between the contributions from short and long trajectories, effectively providing \emph{which-way} information about the path the electron followed during the HHG process. As a result, this prevents the formation of non-classical states such as the one in Eq.~\eqref{Eq:Schrödinger:cat}.

\section{CONCLUSIONS}
In this work, we analyzed the influence of short and long trajectories followed by electrons during the HHG process on the quantum optical state. We observed that these trajectories induce trajectory-dependent displacements on the quantum optical state leading to significant entanglement features between the different harmonic modes in the many-atom regime.~This results in a correlation between the mean photon number of the fundamental mode and the generated harmonic orders.~Heralding operations on specific harmonic orders revealed coherent state superpositions in the fundamental mode, which exhibit negative regions in their Wigner function representation.~To account for the effect of propagation, which spatially separate the contributions of short and long trajectories, we employed an ab-initio propagation model. Our results show that the mean photon number and entanglement features remain like in the case without propagation.~However, the implementation of heralding operations can significantly alter the results, sometimes leading to enhanced entanglement features.~Moreover, we found that physical propagation provides \emph{which-way} information about the electron's path during the HHG process, leading to the absence of non-classical features in the fundamental mode after heralding.

While our analysis has focused on the simpler case where only two trajectories contribute to HHG, based on electron dynamics evaluated within a single laser cycle, in practice HHG processes are driven by multi-cycle laser fields.~This means that, for each laser cycle, multiple pairs of trajectories are involved, with their characteristics influenced by the pulse envelope and carrier-envelope phase.~Consequently, this introduces a diverse range of displacements affecting the final quantum optical state. Future work will explore how these more realistic scenarios could impact the features observed in this study.

Additionally, in solid-state systems, the HHG process involves acceleration steps across different energetic bands of the material~\cite{vampa_theoretical_2014,li_reciprocal-space-trajectory_2019}, similar to atomic systems but constrained by the dispersion relation of each band.~We anticipate that these different propagations will uniquely affect the quantum optical state---as already observed recently~\cite{rivera-dean_nonclassical_2024}---potentially leading to a more intricate entanglement structure that could include matter degrees of freedom given that, unlike in atomic HHG, electrons can end up in various final states~\cite{osika_wannier-bloch_2017,parks_wannier_2020,yue_imperfect_2020}.~Understanding how these results could relate to recent quantum optical treatments of HHG in semiconductors~\cite{gonoskov_nonclassical_2024} and strongly-correlated systems~\cite{lange_electron-correlation-induced_2024}, as well as experimental observations of two-mode squeezing in the low harmonic regime after HHG in semiconductors~\cite{theidel_evidence_2024}, will be an exciting avenue for future research.~

Finally, this work contributes to the expanding field at the intersection of strong-field physics and quantum optics, which in recent years has unveiled phenomena that enhance our understanding of strong-field-driven processes or even present situations that challenge them~\cite{gorlach_high-harmonic_2023,even_tzur_photon-statistics_2023,weber_quantum_2023,stammer_absence_2024,stammer_limitations_2024}.~This, in itself, makes this research direction of great interest, besides the implications the results obtained by the community may have for attosecond science and quantum information science applications~\cite{bhattacharya_stronglaserfield_2023,lewenstein_attosecond_2024,cruz-rodriguez_quantum_2024}.

\section{Acknowledgments}
I acknowledge fruitful discussions with Philipp Stammer, Marcelo F. Ciappina and Maciej Lewenstein.

ICFO QOT group acknowledges support from: European Research Council AdG NOQIA; MCIN/AEI (PGC2018-0910.13039/501100011033, CEX2019-000910-S/10.13039/501100011033, Plan National~FIDEUA PID2019-106901GB-I00, Plan National STAMEENA PID2022-139099NB, I00, project funded by MCIN/AEI/10.13039/501100011033 and by the ``European Union NextGenerationEU/PRTR'' (PRTR-C17.I1), FPI); QUANTERA MAQS PCI2019-111828-2;  QUANTERA DYNAMITE PCI2022-132919, QuantERA II Programme co-funded by European Union’s Horizon 2020 program under Grant Agreement No 101017733; Ministry for Digital Transformation and of Civil Service of the Spanish Government through the QUANTUM ENIA project call - Quantum Spain project, and by the European Union through the Recovery, Transformation and Resilience Plan - NextGenerationEU within the framework of the Digital Spain 2026 Agenda; Fundació Cellex; Fundació Mir-Puig; Generalitat de Catalunya (European Social Fund FEDER and CERCA program, AGAUR Grant No. 2021 SGR 01452, QuantumCAT \ U16-011424, co-funded by ERDF Operational Program of Catalonia 2014-2020); Barcelona Supercomputing Center MareNostrum (FI-2023-3-0024); Funded by the European Union.~Views and opinions expressed are however those of the author(s) only and do not necessarily reflect those of the European Union, European Commission, European Climate, Infrastructure and Environment Executive Agency (CINEA), or any other granting authority.~Neither the European Union nor any granting authority can be held responsible for them (HORIZON-CL4-2022-QUANTUM-02-SGA  PASQuanS2.1, 101113690, EU Horizon 2020 FET-OPEN OPTOlogic, Grant No 899794),  EU Horizon Europe Program (This project has received funding from the European Union’s Horizon Europe research and innovation program under grant agreement No 101080086 NeQSTGrant Agreement 101080086 — NeQST); ICFO Internal ``QuantumGaudi'' project; European Union’s Horizon 2020 program under the Marie Sklodowska-Curie grant agreement No 847648.
\bibliography{References.bib}{}

\clearpage
\onecolumngrid
\appendix

\begin{center}
	\large{\textbf{\textsc{Appendix}}}
\end{center}
\section{Solving the time-dependent Schrödinger equation under the Strong-Field Approximation for a single atom}\label{App:Solution}
In this subsection, we present a detailed derivation of the time-dependent Schrödinger equation (TDSE) describing the dynamics of a single atom embedded in a highly intense electromagnetic field within a quantum optical framework. Under the dipole and single-active electron approximations, and working in the length gauge, the equation can be expressed as
\begin{equation}\label{Eq:App:Or:TDSE}
	i\hbar \pdv{\ket{\Psi(t)}}{t}
		= \Big[
				\hat{H}_{\text{at}}
				+ \mathsf{e}\hat{E} \hat{r}
				+ \hat{H}_{\text{field}}
			\Big] \ket{\Psi(t)},
\end{equation}
where $\hat{H}_{\text{at}}$ denotes the atomic Hamiltonian, and $\mathsf{e}\hat{E}\hat{r}$ represents the light-matter coupling within the considered approximations.~Here $\mathsf{e}$ is the electron's charge, $\hat{E}$ the electric field operator, and $\hat{r}$ is the position operator.~The free-field Hamiltonian is given by $\hat{H}_{\text{field}} = \sum^{q_c}_{q=1} \hbar \omega_q \hat{a}^\dagger_q\hat{a}_q$, where for simplicity the summation runs over a discrete set of field modes.~In the following, we  assume the involved fields to be linearly polarized field, allowing us to effectively restrict the analysis to one dimension. 

When transitioning to the interaction picture with respect to $\hat{H}_{\text{field}}$, the electric field operator becomes time-dependent and is expressed as
\begin{equation}
	\hat{E}(t)
		= -i \sum_{q=1}^{q_c} 
				g(\omega_q)
					\big[
						\hat{a}_q e^{-i\omega_q t} - \hat{a}_q^\dagger e^{i\omega_q t}
					\big],
\end{equation}
where $g(\omega_q)\equiv \sqrt{\hbar\omega_q/(2\epsilon_0 V)}$ is a factor that arises from the expansion of the electric field operator in terms of the creation and annihilation operators. In this picture, where we substitute $\ket{\Psi(t)} = e^{-i H_{\text{field}}t/\hbar}\lvert\Tilde{\Psi}(t)\rangle$, Eq.~\eqref{Eq:App:Or:TDSE} can be written as
\begin{equation}\label{Eq:App:TDSE:2}
	i\hbar \pdv{\lvert\Tilde{\Psi}(t)\rangle}{t}
		= \big[
				\hat{H}_{\text{at}}
				+ \mathsf{e}\hat{E}(t) \hat{r}
			\big]\lvert\Tilde{\Psi}(t)\rangle.
\end{equation}

In the following, we consider \emph{standard} HHG setups, where the electron is initially in its ground state, the single-mode driving field is in a coherent state of amplitude $\alpha$, and all other harmonic modes are in a vacuum state $\ket{0}$. Explicitly, the initial state is given by $\ket{\Psi(t_0)} = \lvert \Tilde{\Psi}(t_0)\rangle = \ket{\text{g}}\bigotimes_{q=1}^{q_c} \ket{\alpha \delta_{q,1}}$.~However, to simplify the analysis and begin with a vacuum state in all modes, we move to a displaced frame of reference $\lvert \Tilde{\Psi}(t) \rangle=\hat{D}_{q=1}(\alpha) \lvert\bar{\Psi}(t)\rangle$, where $\hat{D}_{q}(\cdot)$ is the displacement operator acting on the $q$-th harmonic mode. Substituting this into Eq.~\eqref{Eq:App:TDSE:2} yields
\begin{equation}\label{Eq:App:TSDSE:3}
		i \hbar \pdv{\lvert\bar{\Psi}(t)\rangle}{t}
		= \big[
				\hat{H}_{\text{at}}
				+ \mathsf{e}E_{\text{cl}}(t)\hat{r}
				+ \mathsf{e}\hat{E}(t) \hat{r}
			\big]\lvert\bar{\Psi}(t)\rangle,
\end{equation}
where $E_{\text{cl}}(t) = \text{tr}(\hat{E}(t) \bigotimes^{q_c}_{q=1}\dyad{\alpha\delta_{q,1}})$ represents the classical electric field. In this frame of reference, the initial state becomes $\lvert\bar{\Psi}(t)\rangle = \ket{\text{g}}\otimes\ket{\bar{0}}$ where $\ket{\bar{0}} = \bigotimes_{q=1}^{q_c} \ket{0}$ represents the vacuum state across all modes.

We proceed to solve Eq.~\eqref{Eq:App:TSDSE:3} using an ansatz inspired on the Strong-Field Approximation (SFA), as described in Refs.~\cite{lewenstein_theory_1994,amini_symphony_2019}
\begin{equation}\label{Eq:SFA:ansatz}
	\lvert \bar{\Psi}(t)\rangle
		= a(t) \ket{\text{g}}\otimes \ket{\Phi_{\text{g}}(t)}
			+ \int \dd v \ b(v,t) \ket{v} \otimes \ket{\Phi(v,t)},
\end{equation}
where $\{\ket{v}\}$ denotes the set of atomic scattering states satisfying $\hat{H}_{\text{at}}\ket{v} = v^2/(2m_{\mathsf{e}})\ket{v}$, with $m_{\mathsf{e}}$ being the electron's mass. The states $\ket{\Phi_{\text{g}}(t)}$ and $\ket{\Phi(v,t)}$ represent the quantum optical components when the electron is in the ground state or one of the continuum states, respectively. In the context of the SFA, we assume that (1) all bound states between the ground and continuum states are neglected, and (2) that $\mel{v}{\hat{r}}{v'} \approx i\hbar \pdv*{[ \delta(v-v')]}{v}$, ignoring the contribution of continuum-continuum rescattering events~\cite{lewenstein_theory_1994,amini_symphony_2019}. Under these assumptions, projecting Eq.~\eqref{Eq:App:TSDSE:3} onto the ground and continuum states, while applying the ansatz in Eq.~\eqref{Eq:SFA:ansatz}, respectively yields
\begin{equation}\label{Eq:App:Diff:ground}
	 i \hbar \pdv{}{t}
			\big(
				a(t) \ket{\Phi_{\text{g}}(t)}
			\!\big)
		= -I_p a(t)\ket{\Phi_{\text{g}}(t)}
			+ \mathsf{e} \big( E_{\text{cl}}(t) + \hat{E}(t)\big)
				\int \dd v \mel{\text{g}}{\hat{r}}{v} b(v,t) \ket{\Phi(v,t)},
\end{equation}
\begin{equation}\label{Eq:App:Diff:excited}
		\begin{aligned}
		i\hbar \pdv{}{t}
			\big(
				b(v,t)\ket{\Phi(v,t)}
			\!\big)
			&= \dfrac{v^2}{2\me}b(v,t) \ket{\Phi(v,t)}
				+ \mathsf{e}
				\big(E_{\text{cl}}(t) + \hat{E}(t)\big)
				\mel{v}{\hat{r}}{\text{g}}
				a(t)
				\ket{\Phi_{\text{g}}(t)}
				\\&\quad
					+ i \hbar \mathsf{e}\big(E_{\text{cl}}(t) + \hat{E}(t)\big)
					\pdv{}{v}\big(b(v,t)\ket{\Phi(v,t)}\!\big),
		\end{aligned}
\end{equation}
where in Eq.~\eqref{Eq:App:Diff:ground} we have assumed that the ground state of the atomic system has a well-defined parity, which implies $\mel{\text{g}}{\hat{r}}{\text{g}} = 0$, and used $I_p$ to denote the ionization potential. We will now proceed to solve these equations, starting with the analysis of Eq.~\eqref{Eq:App:Diff:excited}.

\subsection{Solving the equation describing dynamics in the continuum}\label{App:Sol:Exc}
We begin by analyzing Eq.~\eqref{Eq:App:Diff:excited}, which describes the dynamics of the electron in the continuum, and the interactions between the electron and the quantum optical state.~Specifically, examining the right-hand side of the equation, we find that:
\begin{itemize}
	\item The first term represents the energy of the electron in the continuum.
	\item The second term represents the coupling between the ground and continuum states, mediated by the mean value of the applied electric field $E_{\text{cl}}(t)$ and its fluctuations $\hat{E}(t)$.
	\item To interpret the third term, we first decompose it using the chain rule
	\begin{equation}
		\underbrace{\mathsf{e}E_{\text{cl}}(t)\pdv{}{v}\big(b(v,t)\ket{\Phi(v,t)}\!\big)}_{\text{\normalsize \ding{172}}}
		+   \underbrace{\mathsf{e}\hat{E}(t)\pdv{b(v,t)}{v}\ket{\Phi(v,t)}}_{\text{\normalsize\ding{173}}}
		+   \underbrace{\mathsf{e}\hat{E}(t)b(v,t)\pdv{\ket{\Phi(v,t)}}{v}}_{\text{\normalsize\ding{174}}},
	\end{equation}
	where \ding{172} represents a combined effect of the classical field on both the electron dynamics and the quantum optical state, accounting for the consequences of electron acceleration due to the applied electromagnetic field~\cite{lewenstein_theory_1994,amini_symphony_2019}; \ding{173} denotes the back-action of the electron's acceleration on the quantum optical state of the field, similar to a charge current coupling to the electric field operator~\cite{ScullyBookCh2}; and \ding{174} indicates the influence of quantum optical fluctuations on the electron dynamics, illustrating how the quantum optical state, through its fluctuations and coupling to the electron, can modify the electron's trajectory.
\end{itemize}

Regarding \ding{174}, Ref.~\cite{even_tzur_photon-statistics_2023} observed that when non-classical photon statistics are considered as drivers of HHG processes, this back-action becomes significant and non-trivially modifies the electron's trajectories. However, for coherent states drivers, these effects were found to be negligible.~Motivated by this observation, as well as of the great success of semiclassical theories in describing the electron dynamics~\cite{shafir_resolving_2012,pedatzur_attosecond_2015}, we neglect these contributions in the present work.~Consequently, we approximate the differential equation in Eq.~\eqref{Eq:App:Diff:excited} as
\begin{equation}\label{Eq:App:Diff:excited:2}
	\begin{aligned}
		i\hbar \pdv{}{t}
			\big(
				b(v,t)\ket{\Phi(v,t)}
			\big)
			&\approx \dfrac{v^2}{2\me}b(v,t) \ket{\Phi(v,t)}
					+ \mathsf{e}
						\big(E_{\text{cl}}(t) + \hat{E}(t)\big)
						\mel{v}{\hat{r}}{\text{g}}
						a(t)
						\ket{\Phi_{\text{g}}(t)}
					\\&\quad
					+ i \hbar \mathsf{e}E_{\text{cl}}(t)
					\pdv{}{v}\big(b(v,t)\ket{\Phi(v,t)}\!\big)
					+  i \hbar \mathsf{e} \hat{E}(t)
					\pdv{b(v,t)}{v}\ket{\Phi(v,t)}.
	\end{aligned}
\end{equation}

This equation is a first-order differential equation with well-defined homogeneous and inhomogeneous components. Therefore, its solution can be expressed as a combination of the solution to the homogeneous equation and a particular solution to the inhomogeneous part.~Accordingly, we now focus on solving the homogeneous equation---denoted with a subscript ``h'' for clear distinction---which reads as
\begin{equation}
	i\hbar \pdv{}{t}
		\big(
			b_{\text{h}}(v,t)\ket{\Phi_{\text{h}}(v,t)}
		\!\big)
		= \dfrac{v^2}{2\me}b_{\text{h}}(v,t) \ket{\Phi_{\text{h}}(v,t)}
		+ i \hbar \mathsf{e}E_{\text{cl}}(t)
			\pdv{}{v}\big(b_{\text{h}}(v,t)\ket{\Phi_{\text{h}}(v,t)}\!\big)
		+  i \hbar \mathsf{e} \hat{E}(t)
			\pdv{b_{\text{h}}(v,t)}{v}\ket{\Phi_{\text{h}}(v,t)},
\end{equation} 
where we distinguish between two distinct contributions
\begin{equation}\label{Eq:App:homog:excited}
	\begin{aligned}
		&\bigg[
			i\hbar \pdv{b_{\text{h}}(v,t)}{t}
				- \dfrac{v^2}{2\me}b_{\text{h}}(v,t)
				-  i \hbar \mathsf{e}E_{\text{cl}}(t)
					\pdv{b_{\text{h}}(v,t)}{v}
		\bigg] \ket{\Phi_{\text{h}}(v,t)}
		\\
		&\hspace{1cm} 
		+\bigg[
				i\hbar b_{\text{h}}(v,t)
					\pdv{\ket{\Phi_{\text{h}}(v,t)}}{t}
				- i\hbar \mathsf{e} E_{\text{cl}}(t) b_{\text{h}}(v,t) \pdv{\ket{\Phi_{\text{h}}(v,t)}}{t}
				- i\hbar \mathsf{e} \hat{E}(t) 
					\pdv{b_{\text{h}}(v,t)}{v}
						\ket{\Phi_{\text{h}}(v,t)}
			\bigg]
						=0,
	\end{aligned}
\end{equation}
with the first term describing the propagation of an electron in the continuum driven by the classical field $E_{\text{cl}}(t)$. Consequently, the strategy we adopt to solve this differential equation is to find a function $b_{\text{h}}(v,t)$ that makes the first bracket equal to zero, i.e. 
\begin{equation}
		i\hbar \pdv{b_{\text{h}}(v,t)}{t}
		-  i \hbar \mathsf{e}E_{\text{cl}}(t)
			\pdv{b_{\text{h}}(v,t)}{v}
		= \dfrac{v^2}{2\me}b_{\text{h}}(v,t).
\end{equation}

This equation can be easily solved by moving to the canonical momentum frame of reference.~In the context of the SFA, the canonical momentum is a conserved quantity of the electron's motion and is related to the kinetic momentum by $p = m_{\mathsf{e}}v - \mathsf{e}A_{\text{cl}}(t)$, where $A_{\text{cl}}(t)$ is the field's vector potential, related to the electric field by $E_{\text{cl}}(t) = - \pdv*{A_{\text{cl}}(t)}{t}$. Thus, we obtain
\begin{equation}\label{Eq:App:Sol:Excited}
	i\hbar \dv{b_{\text{h}}(v,t)}{t} 
		= \dfrac{\big(p+\mathsf{e}A_{\text{cl}}(t)\big)^2}{2m_{\mathsf{e}}}b_{\text{h}}(v,t)
	\Rightarrow
	b_{\text{h}}(v,t) 
		= b_{\text{h}}(v,t_0)
			\exp[-\dfrac{i}{2m_{\mathsf{e}}\hbar}
					\int^t_{t_0} \dd \tau 
						\big(
							p + \mathsf{e}A_{\text{cl}}(\tau)
						\big)^2],
\end{equation}
and whose partial derivative with respect to $v$ leads to
\begin{equation}\label{Eq:App:displacement}
	\pdv{b_{\text{h}}(v,t)}{v}
		= -\dfrac{i}{\hbar}\Delta r(v,t,t_0) b(v,t)
		 \quad \text{with} \quad 
	\Delta r(v,t,t_0) 
		= \dfrac{1}{m_{\mathsf{e}}}
			\int^{t}_{t_0} \dd \tau 
				\big(
					p + \mathsf{e} A_{\text{cl}}(\tau)
				\big),
\end{equation}
where $\Delta r(v,t,t_0)$ denotes the propagation performed by the electron within the time interval $[t_0,t]$.

By substituting Eqs.~\eqref{Eq:App:Sol:Excited} and \eqref{Eq:App:displacement} into Eq.~\eqref{Eq:App:homog:excited}, we effectively eliminate the first bracket, leaving only the second term. More specifically, by introducing the definition of the canonical momentum, we arrive at
\begin{equation}
	i\hbar \dv{\ket{\Phi_{\text{h}}(v,t)}}{t}
		= \mathsf{e}\hat{E}(t)
			\Delta r(v,t,t_0)\ket{\Phi_{\text{h}}(v,t)},
\end{equation}
that describes the field generated by the charge $\mathsf{e}\Delta r(v,t,t_0)$. Given that this current couples with all field modes simultaneously but independently, the equation results in a displacement operator acting on all modes~\cite{ScullyBookCh2,lewenstein_generation_2021,rivera-dean_light-matter_2022,rivera-dean_strong_2022,stammer_quantum_2023}
\begin{equation}\label{Eq:App:Sol:QO:Excited:hom}
	\ket{\Phi_{\text{h}}(v,t)}
		=
			\hat{\vb{D}}\big(\boldsymbol{\delta}(v,t,t_0)\big)
				\ket{\Phi_{\text{h}}(v,t_0)},
\end{equation}
where we define $\hat{\vb{D}}(\boldsymbol{\delta}(v,t,t_0)) \equiv \bigotimes_{q=1}^{q_c} e^{i\varphi_q(v,t,t_0)} \hat{D}(\delta_q(v,t,t_0))$ with $\delta_q(v,t,t_0)$ and $\varphi_q(v,t,t_0)$ given by
\begin{align}
	& \delta_q(v,t,t_0)
		= \dfrac{\mathsf{e}}{\hbar} g(\omega_q)
			\int_{t_0}^t \dd \tau
				\Delta r(v,t,t_0) e^{i\omega_q \tau},
	\\& \varphi_q(v,t,t_0)
		= \dfrac{\mathsf{e}^2}{\hbar^2} g(\omega_q)^2
			\int_{t_0}^t \dd \tau_2 \int_{t_0}^{t_2} \dd \tau_1
				\Delta r(v,t_2,t_0)\Delta r(v,t_1,t_0) \sin(\omega_q(t_2-t_1)).
\end{align}

Consequently, by combining Eqs.~\eqref{Eq:App:Sol:Excited} and \eqref{Eq:App:Sol:QO:Excited:hom}, the solution to the homogeneous equation is given by
\begin{equation}
	b_{\text{h}}(v,t)\ket{\Phi_{\text{h}}(v,t)}
		= b_{\text{h}}(v,t_0)\exp[-\dfrac{i}{2m_{\mathsf{e}}\hbar}
					\int^t_{t_0} \dd \tau 
					\big(
						p + \mathsf{e}A_{\text{cl}}(\tau)
					\big)^2]
			\hat{\vb{D}}\big(\boldsymbol{\delta}(v,t,t_0)\big)
			\ket{\Phi_{\text{h}}(v,t_0)},
\end{equation}
such that a solution for the first-order inhomogeneous differential equation, once the initial conditions are introduced, is given by
\begin{equation}\label{Eq:App:Final:Sol:Excited}
	b(v,t)\ket{\Phi(v,t)}
		= -\dfrac{i}{\hbar}
			\int^t_{t_0}
				\dd t_1
					e^{-iS(p,t,t_1)/\hbar}
					\hat{\vb{D}}\big(\boldsymbol{\delta}(v,t,t_1)\big)
					\big(
						E_{\text{cl}}(t_1) + \hat{E}(t_1)
					\big)
					\mel{p + \mathsf{e}A_{\text{cl}}(t_1)}{\hat{r}}{\text{g}}
					a(t_1)\ket{\Phi_{\text{g}}(t_1)},
\end{equation}
where $S(p,t,t_1) \equiv [1/(2m_{\mathsf{e}})] \int^t_{t_1} \dd \tau (p+\mathsf{e}A_{\text{cl}}(\tau))^2$. We note that, in the man text, the contributions of $\varphi_q(v,t,t_0)$ are not considered for being proportional to $g(\omega_q)^2$.

\subsection{Solving the equation describing dynamics in the ground state}\label{App:Sol:ground}
Now, we turn our attention to Eq.~\eqref{Eq:App:Diff:ground}, which is also a first-order differential equation with well-defined homogeneous and inhomogeneous components. For the homogeneous part, we have
\begin{equation}
	i\hbar \pdv{\big(a_{\text{h}}(t)\ket{\Phi_{\text{g,\text{h}}}(t)}\!\big)}{t}
		= -I_p a_{\text{h}}(t)\ket{\Phi_{\text{g},\text{h}}(t)}
	\Rightarrow
	a_{\text{h}}(t)\ket{\Phi_{\text{g},\text{h}}(t)}
		= e^{iI_pt/\hbar}a_\text{h}(t_0)\ket{\Phi_{\text{g},\text{h}}(t_0)},
\end{equation}
such that a solution to Eq.~\eqref{Eq:App:Diff:ground}, considering the initial conditions, can be written as
\begin{equation}\label{Eq:App:Sol:ground}
	\begin{aligned}
	a(t)\ket{\Phi_{\text{g}}(t)}
		&= e^{iI_p (t-t_0)/\hbar} \ket{\bar{0}}
			\\&\quad
			 - \dfrac{1}{\hbar}^2
			\int_{t_0}^t \dd t_2 \int \dd p \int^{t_2}_{t_0} \dd t_1
				e^{iI_p(t-t_2)/\hbar}  \big(E_{\text{cl}}(t_2) + \hat{E}(t_2)\big)
				 \mel{\text{g}}{\hat{r}}{p+\mathsf{e}A_{\text{cl}}(t_2)}
					e^{-iS(p,t_2,t_1)/\hbar}
			\\&\hspace{4.3cm}\times
					\hat{\vb{D}}\big(\boldsymbol{\delta}(p,t_2,t_1)\big)		
					\big(E_{\text{cl}}(t_1) + \hat{E}(t_1)\big)
					\mel{p+\mathsf{e}A_{\text{cl}}(t_1)}{\hat{r}}{\text{g}}
					a(t_1)\ket{\Phi_{\text{g}}(t_1)},
	\end{aligned}
\end{equation}
where Eq.~\eqref{Eq:App:Final:Sol:Excited} has been explicitly introduced, and the result is expressed in terms of the canonical momentum $p$ for convenience.~As noted, this leads to a self-recursive expression reminiscent of a Dyson expansion, where a single iteration suffices to recover the HHG dynamics. Specifically, we obtain
\begin{equation}\label{Eq:App:Sol:ground:approx}
	\begin{aligned}
		a(t)\ket{\Phi_{\text{g}}(t)}
			&\approx e^{iI_p (t-t_0)/\hbar} \ket{\bar{0}}
			\\&\quad
				- \dfrac{1}{\hbar}^2
					\int_{t_0}^t \dd t_2 \int \dd p \int^{t_2}_{t_0} \dd t_1
						e^{iI_p(t-t_2)/\hbar} \big(E_{\text{cl}}(t_2) + \hat{E}(t_2)\big)
						\mel{\text{g}}{\hat{r}}{p+\mathsf{e}A_{\text{cl}}(t_2)}
						e^{-iS(p,t_2,t_1)/\hbar}
						\\&\hspace{4.3cm}\times		
							\hat{\vb{D}}\big(\boldsymbol{\delta}(p,t_2,t_1)\big)
							\big(E_{\text{cl}}(t_1) + \hat{E}(t_1)\big)
							\mel{p+\mathsf{e}A_{\text{cl}}(t_1)}{\hat{r}}{\text{g}}
							e^{iI_p (t_1-t_0)/\hbar} \ket{\bar{0}},
	\end{aligned}
\end{equation}
where the first term on the right hand side represents the scenario where the electron does not interact with the field and remains in the ground state.~In contrast, the second term on the right-hand side reveals the three-step mechanism leading to HHG and the resulting effects on the quantum optical state:
\begin{enumerate}
	\item \textbf{Ionization.}~At time $t_1$, the light-matter interaction causes the electron to transition from the ground state to the continuum state $\ket{p+\mathsf{e}A_{\text{cl}}(t)}$, driven by the classical electromagnetic field $E_{\text{cl}}(t)$ and its quantum fluctuations $\hat{E}(t)$.
	\item \textbf{Acceleration.}~Between times $t_1$ and $t_2$ ($t_2 \geq t_1$), the electron accelerates in the continuum due to the field, reaching the state $\ket{p+\mathsf{e}A_{\text{cl}}(t_2)}$ and acquiring an additional phase $S(p,t_2,t_1)$.~This acceleration results in a charge current that couples to the electromagnetic field, causing a displacement $\boldsymbol{\delta}(p,t_2,t_1)$ in the quantum optical state, which depends on the ionization conditions $(p,t_1)$~\cite{rivera-dean_light-matter_2022} and the evolution up to $t_2$.
	\item \textbf{Recombination.}~At time $t_2$, the electron transitions from the continuum state $\ket{p+\mathsf{e}A_{\text{cl}}(t_2)}$ back to the ground state, with this transition coupled to the electromagnetic field (both classical and quantum fluctuations contribution). Once in the ground state, the electron acquires an additional phase $I_p(t-t_2)/\hbar$ due to its evolution in the ground state until the final time $t$.
\end{enumerate}

It is important to note that the modifications in the quantum optical state of the field due to the electron dynamics in the continuum do not affect the electronic trajectories, which thus align with the semiclassical predictions~\cite{lewenstein_theory_1994,amini_symphony_2019}. This is due to the neglect of the contribution from \ding{174} in Eq.~\eqref{Eq:App:Diff:excited}, which is expected to be negligible for coherent state drivers at laser intensities that do not significantly deplete the driving field.~Furthermore, the additional terms in Eq.~\eqref{Eq:App:Sol:ground} that appear beyond a single recursive iteration represent scenarios where the same electron undergoes HHG dynamics multiple times, though with decreasing probability as the number of iterations increases.~For this work, we focus on the dynamics happening within a single cycle of the field unless otherwise specified, and therefore restrict our analysis to Eq.~\eqref{Eq:App:Sol:ground:approx}.

\subsection{The (approximated) final state of the joint system}\label{App:Final:state}
By substituting Eqs.~\eqref{Eq:App:Sol:Excited} and \eqref{Eq:App:Sol:ground} in our ansatz, and considering a single recursive iteration as outlined in Eq.~\eqref{Eq:App:Sol:ground:approx}, we can express the final state of the joint electron-field system as
\begin{equation}\label{Eq:App:Final:state}
	\lvert\bar{\Psi}(t)\rangle
		\approx 
			\lvert \bar{\Psi}_0(t)\rangle 
			+ \lvert \bar{\Psi}_{\text{ATI}}(t)\rangle
			+ \lvert \bar{\Psi}_{\text{HHG}}(t)\rangle,
\end{equation}
where $\lvert \bar{\Psi}_0(t)\rangle = e^{iI_p(t-t_0)/\hbar}\ket{\text{g}}\otimes \ket{\bar{0}}$ represents the quantum state of the system if no interaction has occurred, while the other two contributions correspond to direct ATI events~\cite{rivera-dean_light-matter_2022} and HHG events, respectively. More explicitly, these two contributions are given by
\begin{equation}
	\lvert\bar{\Psi}_{\text{ATI}}(t)\rangle
		= -\dfrac{i}{\hbar} \int \dd p \int^{t}_{t_0} \dd t_1
				e^{-iS(p,t,t_1)/\hbar}
				\hat{\vb{D}}\big(\boldsymbol{\delta}(p,t,t_1)\big)
				\big(
				E_{\text{cl}}(t_1) + \hat{E}(t_1)
				\big)
				\mel{p + \mathsf{e}A_{\text{cl}}(t_1)}{\hat{r}}{\text{g}}
				e^{i I_p(t_1-t_0)/\hbar} \ket{p+\mathsf{e}A_{\text{cl}}(t)}\otimes \ket{\bar{0}},
\end{equation}
\begin{equation}\label{Eq:App:HHG:state}
	\begin{aligned}
		\lvert\bar{\Psi}_{\text{HHG}}(t)\rangle
			&= - \dfrac{1}{\hbar^2}
				\int_{t_0}^t \dd t_2 \int \dd p \int^{t_2}_{t_0} \dd t_1
					e^{iI_p(t-t_2)/\hbar}\hat{\vb{D}}\big(\boldsymbol{\delta}(p,t_2,t_1)\big)
					 \big(E_{\text{cl},\boldsymbol{\delta}}(t_2) + \hat{E}(t_2)\big)
					\mel{\text{g}}{\hat{r}}{p+\mathsf{e}A_{\text{cl}}(t_2)}
					e^{-iS(p,t_2,t_1)/\hbar}
					\\&\hspace{4.3cm}\times		
						\big(E_{\text{cl}}(t_1) + \hat{E}(t_1)\big)
						\mel{p+\mathsf{e}A_{\text{cl}}(t_1)}{\hat{r}}{\text{g}}
						e^{iI_p (t_1-t_0)/\hbar}
						\ket{\text{g}}\otimes \ket{\bar{0}},
	\end{aligned}
\end{equation}
where in Eq.~\eqref{Eq:App:HHG:state} we have moved the displacement operator in front of the expression, and defined $E_{\text{cl},\boldsymbol{\delta}}(t) = E_{\text{cl}}(t) + \Tr(\hat{E}(t)\bigotimes_{q=1}^{q_c}\dyad{\delta_q(p,t_2,t_1)})$.

The state in Eq.~\eqref{Eq:App:Final:state} is generally an entangled state as both atomic and optical degrees of freedom are non-trivially affected, with the properties of the latter depending on the dynamics experienced by the former.~In Ref.~\cite{rivera-dean_light-matter_2022}, the entanglement properties of the direct ATI contribution were analyzed separately.~Here, we focus on the properties of the quantum optical state resulting from events where the electron ends up in the ground state after the interaction, specifically $\ket{\Phi_{\text{g}}(t)}$.~This state is obtained by projecting Eq.~\eqref{Eq:App:Final:state} onto the atomic ground state
\begin{equation}
	\begin{aligned}
	\ket{\Phi_{\text{g}}(t)}
		= \langle\text{g}\vert\bar{\Psi}(t)\rangle
		&\approx 
			\langle \text{g} \vert \bar{\Psi}_0(t)\rangle 
			+\langle \text{g} \vert \bar{\Psi}_{\text{HHG}}(t)\rangle
			\equiv
			\ket{\Phi_0(t)} + \ket{\Phi_{\text{HHG}}(t)},
	\end{aligned}
\end{equation}
where we have denoted $\ket{\Phi_0(t)} \equiv \langle \text{g} \vert \bar{\Psi}_0(t)\rangle$ and $\ket{\Phi_{\text{HHG}}(t)} \equiv 
\langle \text{g} \vert \bar{\Psi}_{\text{HHG}}(t)\rangle$.

By examining $\ket{\Phi_{\text{HHG}(t)}}$ (see Eq.~\eqref{Eq:App:HHG:state}), we can observe different contributions by further expanding the product $(E_{\text{cl}}(t_2) + \hat{E}(t_2))(E_{\text{cl}}(t_1) + \hat{E}(t_1))$. Specifically, we obtain
\begin{equation}
	(E_{\text{cl},\boldsymbol{\delta}}(t_2) + \hat{E}(t_2))(E_{\text{cl}}(t_1) + \hat{E}(t_1))
		= E_{\text{cl},\boldsymbol{\delta}}(t_2)E_{\text{cl}}(t_1)
			+ E_{\text{cl},\boldsymbol{\delta}}(t_2)\hat{E}(t_1)
			+ \hat{E}(t_2) E_{\text{cl}}(t_1)
			+ \hat{E}(t_2)\hat{E}(t_1),
\end{equation}
where the first one does not contribute to the harmonic generation process, as it does not involve the action of creation and annihilation operators on the field modes, while the remaining terms do.~Specifically, the second and third terms, which are proportional to $g(\omega_L)$, lead to creation and annihilation of photons. The fourth terms, which is proportional to $g(\omega_L)^2$, corresponds to second-order effects that could involve squeezing of the field modes~\cite{stammer_squeezing_2023}. 

In the following, we neglect the contributions from second and higher-order terms with respect to $g(\omega_L)$. We also encapsulate the $E_{\text{cl},\boldsymbol{\delta}}(t_2)E_{\text{cl}}(t_1)$ within the $\ket{\Phi_0(t)}$ contribution, though their effect in this analysis is not considered provided that they account for the atomic ground state depletion, which is negligible for the field strengths used in this analysis~\cite{delone_tunneling_1998,tong_empirical_2005}.~Consequently, the contribution of $\ket{\Phi_{\text{HHG}}(t)}$ is expressed as
\begin{equation}
	\begin{aligned}
	\ket{\Phi_{\text{HHG}}(t)}
		&\approx - \dfrac{1}{\hbar^2}
				\int_{t_0}^t \dd t_2 \int \dd p \int^{t_2}_{t_0} \dd t_1
					e^{iI_p(t-t_2)/\hbar}\hat{\vb{D}}\big(\boldsymbol{\delta}(p,t_2,t_1)\big)
					 E_{\text{cl},\boldsymbol{\delta}}(t_2)
					\mel{\text{g}}{\hat{r}}{p+\mathsf{e}A_{\text{cl}}(t_2)}
				e^{-iS(p,t_2,t_1)/\hbar}
				\\&\hspace{4.3cm}\times		
				 \hat{E}(t_1)
				\mel{p+\mathsf{e}A_{\text{cl}}(t_1)}{\hat{r}}{\text{g}}
				e^{iI_p (t_1-t_0)/\hbar}
		\ket{\text{g}}\otimes \ket{\bar{0}}
		\\&\quad 
		- \dfrac{1}{\hbar^2}
				\int_{t_0}^t \dd t_2 \int \dd p \int^{t_2}_{t_0} \dd t_1
					e^{iI_p(t-t_2)/\hbar}\hat{\vb{D}}\big(\boldsymbol{\delta}(p,t_2,t_1)\big)
				\hat{E}(t_2)
		\mel{\text{g}}{\hat{r}}{p+\mathsf{e}A_{\text{cl}}(t_2)}
		e^{-iS(p,t_2,t_1)/\hbar}
		\\&\hspace{4.3cm}\times		
		E_{\text{cl}}(t_1)
		\mel{p+\mathsf{e}A_{\text{cl}}(t_1)}{\hat{r}}{\text{g}}
		e^{iI_p (t_1-t_0)/\hbar}
		\ket{\text{g}}\otimes \ket{\bar{0}}
	\end{aligned},
\end{equation}
and provided that $\hat{E}(t) \ket{0} = i\sum_{q=1}^q g(\omega_q)e^{i\omega_q t} \ket{1}_q \bigotimes^{q_c}_{q'\neq q}\ket{0}$, we can express the equation before as 
\begin{equation}\label{Eq:App:HHG:explicit}
	\begin{aligned}
		\ket{\Phi_{\text{HHG}}(t)}
			&= - \dfrac{1}{\hbar^2}\sum_{q=1}^{q} g(\omega_q)
				\int_{t_0}^t \dd t_2 \int \dd p \int^{t_2}_{t_0} \dd t_1
				e^{iI_p(t-t_2)/\hbar}\hat{\vb{D}}\big(\boldsymbol{\delta}(p,t_2,t_1)\big)
				E_{\text{cl},\boldsymbol{\delta}}(t_2)
				\mel{\text{g}}{\hat{r}}{p+\mathsf{e}A_{\text{cl}}(t_2)}
				e^{-iS(p,t_2,t_1)/\hbar}
				\\&\hspace{5.67cm}\times		
					e^{i\omega_q t_1}
					\mel{p+\mathsf{e}A_{\text{cl}}(t_1)}{\hat{r}}{\text{g}}
					e^{iI_p (t_1-t_0)/\hbar}
					\ket{1}_q \bigotimes^{q_c}_{q'\neq q}\ket{0}
		\\&\quad 
			- \dfrac{1}{\hbar^2}\sum_{q=1}^{q} g(\omega_q)
			\int_{t_0}^t \dd t_2 \int \dd p \int^{t_2}_{t_0} \dd t_1
			e^{iI_p(t-t_2)/\hbar}\hat{\vb{D}}\big(\boldsymbol{\delta}(p,t_2,t_1)\big)
			e^{i\omega_q t_2}
			\mel{\text{g}}{\hat{r}}{p+\mathsf{e}A_{\text{cl}}(t_2)}
			e^{-iS(p,t_2,t_1)/\hbar}
			\\&\hspace{5.67cm}\times
			E_{\text{cl}}(t_1)
			\mel{p+\mathsf{e}A_{\text{cl}}(t_1)}{\hat{r}}{\text{g}}
			e^{iI_p (t_1-t_0)/\hbar}
			\ket{1}_q \bigotimes^{q_c}_{q'\neq q}\ket{0},
	\end{aligned}
\end{equation}
where the main difference between these two terms lies in the timing of photon generation: either at time $t_1$ (after ionization) or at time $t_2$ (after recombination). The latter case corresponds to the three-step HHG model.

\begin{figure}
	\centering
	\includegraphics[width = 0.8\columnwidth]{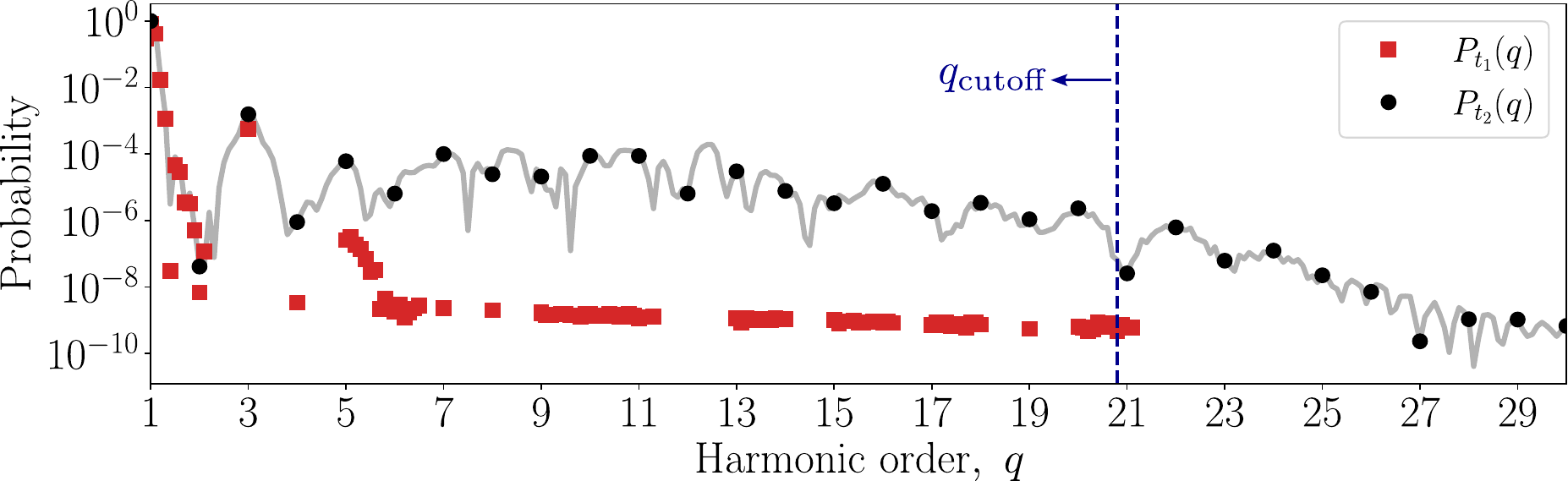}
	\caption{Probability of generating a single photon in a given harmonic mode for the two channels described in Eq.~\eqref{Eq:App:HHG:explicit}: generation of radiation at ionization (red square markers) or at recombination (black circular markers).~The analysis uses an 8 cycle laser pulse with $\sin^2$-envelope, field strength $E_0 = 0.053$ a.u., central frequency $\omega_L = 0.057$ a.u.~and duration $\Delta t \approx 21$ fs was used.~The $y$-axis is normalized to the maximum probability observed.}
	\label{Fig:Comp:Probs}
\end{figure}

To evaluate which term contributes the most in Eq.~\eqref{Eq:App:HHG:explicit} by explicitly calculating the probability of generating a photon in a given harmonic order $q$. This probability is given by
\begin{equation}
	P(q) = \abs{\braket{1_q}{\Phi_{\text{HHG}}(t)}}^2
			= P_{t_1}(q) + P_{t_2}(q) + \text{interference},
\end{equation}
where $P_{t_i}(q)$ represents the probability of generating a harmonic photon at $t_i$.~We separately evaluate $P_{t_1}(q)$ and $P_{t_2}(q)$, and compare their relative contribution.~This comparison is illustrated in Fig.~\ref{Fig:Comp:Probs} for a laser pulse with a sin$^2$-envelope, field strength $E_0 = 0.053$ a.u., central frequency $\omega_L = 0.057$ a.u. and a duration of 8 cycles ($\Delta t \approx 21$ fs).~As shown, $P_{t_2}(q)$ dominates over $P_{t_1}(q)$, particularly in the high-order harmonic regime, indicating that the standard-HHG channel is the dominant contribution.~Therefore, for our analysis, which focuses on the $q\geq 15$ region where the saddle-point approximation becomes applicable, we approximate the HHG state by
\begin{equation}\label{Eq:App:HHG:explicit:approx}
	\begin{aligned}
		\ket{\Phi_{\text{HHG}}(t)}
		&\approx- \dfrac{1}{\hbar^2}\sum_{q=1}^{q} g(\omega_q)
		\int_{t_0}^t \dd t_2 \int \dd p \int^{t_2}_{t_0} \dd t_1
		e^{iI_p(t-t_2)/\hbar}\hat{\vb{D}}\big(\boldsymbol{\delta}(p,t_2,t_1)\big)
		e^{i\omega_q t_2}
		\mel{\text{g}}{\hat{r}}{p+\mathsf{e}A_{\text{cl}}(t_2)}
		e^{-iS(p,t_2,t_1)/\hbar}
		\\&\hspace{5.67cm}\times		
		E_{\text{cl}}(t_1)
		\mel{p+\mathsf{e}A_{\text{cl}}(t_1)}{\hat{r}}{\text{g}}
		e^{iI_p (t_1-t_0)/\hbar}
		\ket{1_q} \bigotimes^{q_c}_{q'\neq q}\ket{0}.
	\end{aligned}
\end{equation}

\section{About saddle-point approximation}\label{App:Saddle:Point}
The saddle-point methods is a mathematical tool used to simplify the evaluation of integrals involving rapidly oscillating functions by approximating them as a sum over carefully chosen points, known as saddle-points. This approach is commonly employed in the analysis of strong-field processes, where the generation of high-order harmonics in strong-laser fields results in rapidly oscillating integrands. This is particularly the case for Eq.~\eqref{Eq:App:HHG:explicit:approx}, which contains complex exponential terms with coefficients that lead to fast oscillations. 

Unlike in traditional semiclassical analyses, in our case the integrals involve operators that explicitly depend on the integration variables.~This could introduce additional phase components when these operators act on specific quantum states, and that might non-trivially modify the location of the saddle-points compared to those obtained in semiclassical analyses.~Nevertheless, we argue that, in the specific scenario considered in this work, where $\abs{\boldsymbol{\delta}} \ll 1$, the saddle-point approximation can still be directly applied to Eq.~\eqref{Eq:App:HHG:explicit:approx} using the same saddle-points as those obtained in semiclassical HHG analyses. 

To demonstrate our point, we introduce a simplified notation for Eq.~\eqref{Eq:App:HHG:explicit:approx} to facilitate analysis, and represent it as
\begin{equation}
	\ket{\Phi_{\text{HHG}}(t)}
		= \int \dd \boldsymbol{\theta}\ M(\boldsymbol{\theta}) 
				D\big(\delta(\boldsymbol{\theta})\big)
				\ket{\phi},
\end{equation}
where we have encapsulated the integration variables into a multi-dimensional vector, $\boldsymbol{\theta}$, and all functions that depend on these variables into $M(\boldsymbol{\theta})$,  except for the displacement operator. For simplicity, we consider the use of a single mode, although the results can be generalized to an arbitrary number of modes.

 After projecting our state onto an integration-variable-independent basis set, such as the Fock basis, we obtain
 \begin{equation}
 	\ket{\Phi_{\text{HHG}}(t)}
 		= \sum_{n=0}^\infty
 				\int \dd \boldsymbol{\theta} \ M(\boldsymbol{\theta})
 		 		\mel{n}{\hat{D}\big(\delta(\boldsymbol{\theta})\big)}{\phi}
 		 		\ket{n}, 
 \end{equation}
where we can distinguish between two scenarios based on the form of $\ket{\phi}$ in Eq.~\eqref{Eq:App:HHG:explicit}. If $\ket{\phi} = \ket{0}$, expanding the matrix element of the displacement operator gives
\begin{equation}
		\ket{\Phi_{\text{HHG}}(t)}
			= \sum_{n=0}^\infty
				\int \dd \boldsymbol{\theta}\ M(\boldsymbol{\theta})
				\dfrac{\delta(\boldsymbol{\theta})^n}{\sqrt{n!}}e^{-|\delta(\boldsymbol{\theta})|^2/2}
				\ket{n},
\end{equation}
while for the case $\ket{\phi}=\ket{1}$ we have
 \begin{equation}
 	\begin{aligned}
	\ket{\Phi_{\text{HHG}}(t)}
		&= \sum_{n=0}^\infty
			\int \dd \boldsymbol{\theta}\ M(\boldsymbol{\theta})
			\bra{n}
				\big(
					\hat{a}^\dagger-\delta(\boldsymbol{\theta})^*
				\big)
				\hat{D}\big(\delta(\boldsymbol{\theta})\big)\ket{0}
			\ket{n}
		\\&=  \sum_{n=0}^\infty
			\bigg[
			\int \dd \boldsymbol{\theta} \ M(\boldsymbol{\theta})
				\sqrt{n}\dfrac{\delta(\boldsymbol{\theta})^{n-1}}{\sqrt{(n-1)!}}e^{-|\delta(\boldsymbol{\theta})|^2/2}\ket{n}
			- \int \dd \boldsymbol{\theta}\ M(\boldsymbol{\theta})
					\delta(\boldsymbol{\theta})^*\dfrac{\delta(\boldsymbol{\theta})^n}{\sqrt{n!}}
					e^{-|\delta(\boldsymbol{\theta})|^2/2}
					\ket{n}
			\bigg].
	\end{aligned}
\end{equation}

\begin{figure}
	\centering
	\includegraphics[width = 0.7\textwidth]{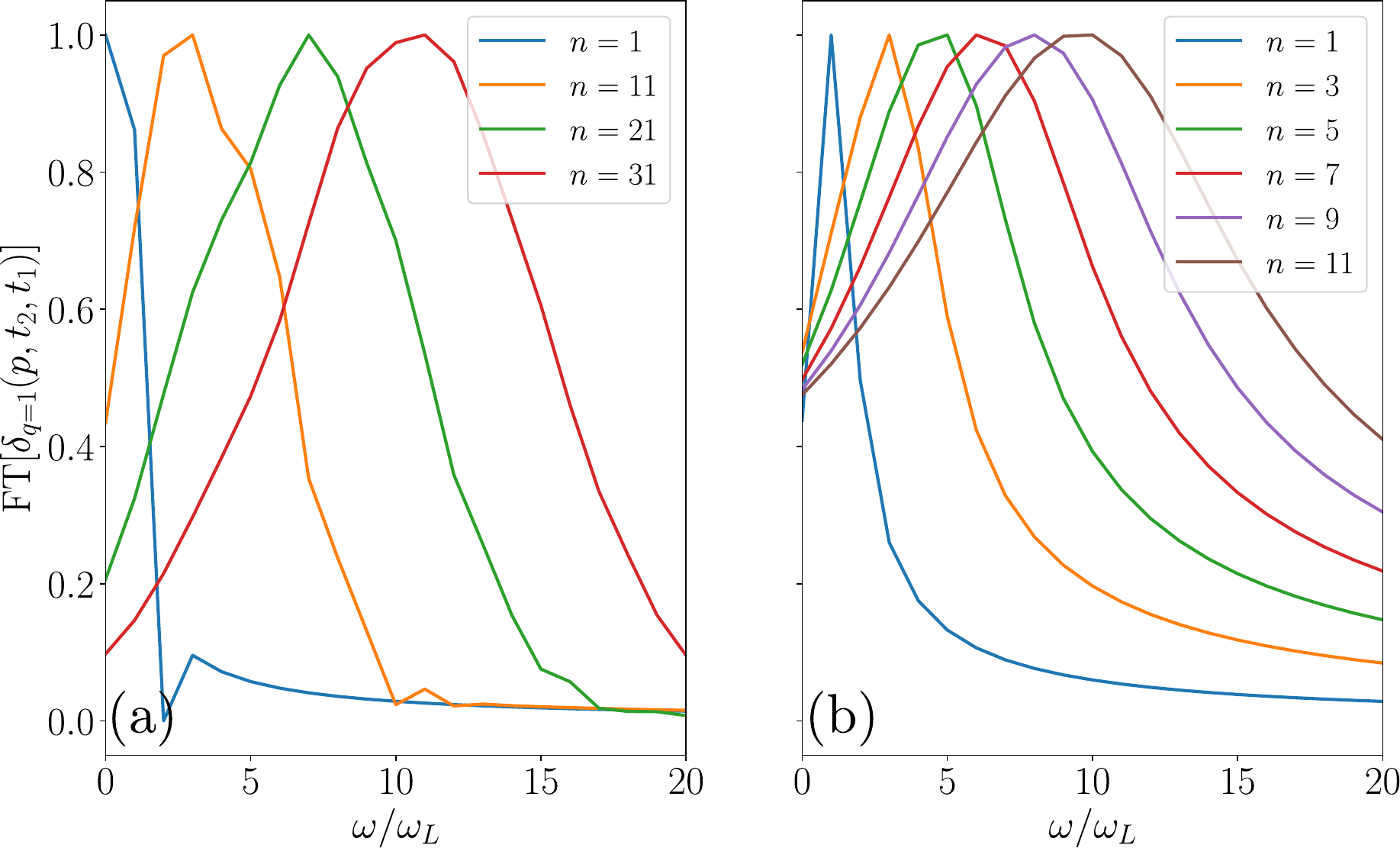}
	\caption{Fourier transform of $\delta(\boldsymbol{\theta})$ for the fundamental mode, considering different powers $n$. The integration limits are set to 0 and $2\pi/\omega_L$.Panel (a) shows results for $p = 0$ a.u., while panel (b) shows results for $ p = 1$ a.u., representing the boundaries of the values obtained when performing the saddle-point approximation.}
	\label{Fig:App:FT:disp}
\end{figure}

In both cases, we observe that powers of $\delta(\boldsymbol{\theta})$ contribute to the integral. If $\delta(\boldsymbol{\theta})$ itself is an oscillating function of the integration variables, their increasingly powers could lead to fast-oscillating contributions, significantly affecting the applicability of the saddle-point approximation as typically used in strong-field contexts.

In Fig.~\ref{Fig:App:FT:disp},  we present the Fourier transform of different powers of $\delta(\boldsymbol{\theta})$ when considering $q=1$, which is the value that contributes the most, as discussed in Fig.~\ref{Fig:Disp:q1:q2} of the main text.~Here, we set $E_0 = 0.053$ a.u.~and $\omega_L = 0.057$ a.u., with integration limits from $t_0=0$ to $t = 2\pi/\omega_L$.~We used $p=0$ a.u.~in panel (a), while $p=1$ a.u.~in panel (b). These values for $p$ were chosen based on those obtained from the semiclassical saddle-point approximation for the field amplitudes and frequencies considered in this work.

As observed, for low values of $n$, the function $\delta(\boldsymbol{\theta})$ oscillates with frequency $\omega_L$ (panel (b)) or less (panel (a)), and thus does not significantly contribute to the highly oscillating phases provided by $M(\boldsymbol{\theta})$.~However, as $n$ increases, the oscillation frequency also increases, reaching values that could significantly impact the semiclassical saddle-point equations ($\omega/\omega_L \approx 15$). Nevertheless, since the single-atom contributions to the amplitude $\lvert\delta(\boldsymbol{\theta})\rvert$ scale with $g(\omega_q)$---which in our case is a perturbative quantity---higher values of $n$ lead to exponentially decreasing contributions of $\lvert\delta(\boldsymbol{\theta})\rvert$. 

Thus, for the scenario considered in this work, we can truncate to low values of $n$ and apply the saddle-point approximation only to the phases arising from $M(\boldsymbol{\theta})$. Consequently, we can approximate Eq.~\eqref{Eq:App:HHG:explicit:approx} as
\begin{equation}
	\ket{\Phi_{\text{HHG}}(t)}
		\approx
			-\dfrac{e^{iI_p t}}{\hbar^2}\sum^q_{q=1}g(\omega_q)
				\sum_{\{t_{\text{re}},p_{\text{s}},t_{\text{ion}}\}}
				\hspace{-0.4cm}G(t_{\text{re}},p_{\text{s}},t_{\text{ion}})
				\hat{\vb{D}}
					\big(
						\boldsymbol{\delta}(p_s,t_{\text{re}},t_{\text{ion}})
					\big)
				M(p_s,t_{\text{re}},t_{\text{ion}})
				\ket{1_q}
				\bigotimes^{q_c}_{q'\neq q}\ket{0},		
\end{equation}
where $\{t_{\text{re}},p_s,t_{\text{ion}}\}$ represents the set of semiclassical saddle-points, $G(t_{\text{re}},p_{\text{s}},t_{\text{ion}})$ contains some prefactors obtained from applying the saddle-point method~\cite{lewenstein_theory_1994,amini_symphony_2019}, and $M(p_s,t_{\text{re}},t_{\text{ion}})$ represents the probability amplitude of an electron undergoing an HHG process defined by the set of variables $(t_{\text{re}},p_s,t_{\text{ion}})$. More explicitly, the latter is given by
\begin{equation}\label{Eq:App:HHG:explicit:saddles}
	M(p_s,t_{\text{re}},t_{\text{ion}})
		\equiv e^{-iS^{(q)}_{\text{sc}}(p_s,t_{\text{re}},t_{\text{ion}})/\hbar}
					\mel{\text{g}}{\hat{r}}{p+\mathsf{e}A_{\text{cl}}(t_{\text{re}})}
					E_{\text{cl}}(t_{\text{ion}})
					\mel{p+\mathsf{e}A_{\text{cl}}(t_{\text{ion}})}{\hat{r}}{\text{g}},
\end{equation}
where $S^{(q)}_{\text{sc}}(p_s,t_{\text{re}},t_{\text{ion}}) \equiv S(p_s,t_{\text{re}},t_{\text{ion}}) + I_p(t_{\text{re}}-t_{\text{ion}}) - q\hbar\omega_Lt_{\text{re}}$ is the semiclassical HHG action.

\subsection{The semiclassical saddle-point equations and its numerical evaluation}
The saddle-points over which Eq.~\eqref{Eq:App:HHG:explicit:saddles} is defined are obtained by looking for the saddle-points of the semiclassical HHG action $S^{(q)}_{\text{sc}}(p,t_2,t_1)$. These can be obtained by setting the partial derivatives of this function with respect to each of the integration variables, more explicitly
\begin{align}
	& \pdv{S^{(q)}_{\text{sc}}(p,t_2,t_1)}{t_1}\bigg\rvert_{\substack{p_s, t_{\text{re}},\\ t_{\text{ion}}}} = 0 
		\Rightarrow \dfrac{\big[p + A_{\text{cl}}(t_{\text{ion}})\big]^2}{2m_{\mathsf{e}}}
								+ I_p = 0,\label{Eq:App:Ionization}
	\\
	& \pdv{S^{(q)}_{\text{sc}}(p,t_2,t_1)}{p}\bigg\rvert_{\substack{p_s, t_{\text{re}},\\ t_{\text{ion}}}} = 0  
		\Rightarrow \int^{t_{\text{re}}}_{t_{\text{ion}}} \dd \tau
			\big[p_s + A_{\text{cl}}(\tau)\big] = 0,
	\\
	& \pdv{S^{(q)}_{\text{sc}}(p,t_2,t_1)}{t_2}\bigg\rvert_{\substack{p_s, t_{\text{re}},\\ t_{\text{ion}}}} = 0 
		\Rightarrow
			\dfrac{\big[p + A_{\text{cl}}(t_{\text{re}})\big]^2}{2m_{\mathsf{e}}}
			+ I_p = \hbar q\omega_L,\label{Eq:App:Recombination}
\end{align}
which describe the ionization, acceleration and recombination steps described earlier, respectively.

It is worth noting that, for these equations to be solvable, the use of complex-valued parameters is needed which account for the nuances about the tunneling processes leading to HHG. In our case, we solved these equations numerically in \texttt{Mathematica} using the function \texttt{FindComplexRoots} of the package \texttt{EPToolbox}~\cite{pisanty_episantyeptoolbox_2018}. In order to ensure convergence of the employed method, we considered a total number of 1000 initial seeds and set the maximum number of iterations to 1000.

\begin{figure}
	\centering
	\includegraphics[width=0.9\textwidth]{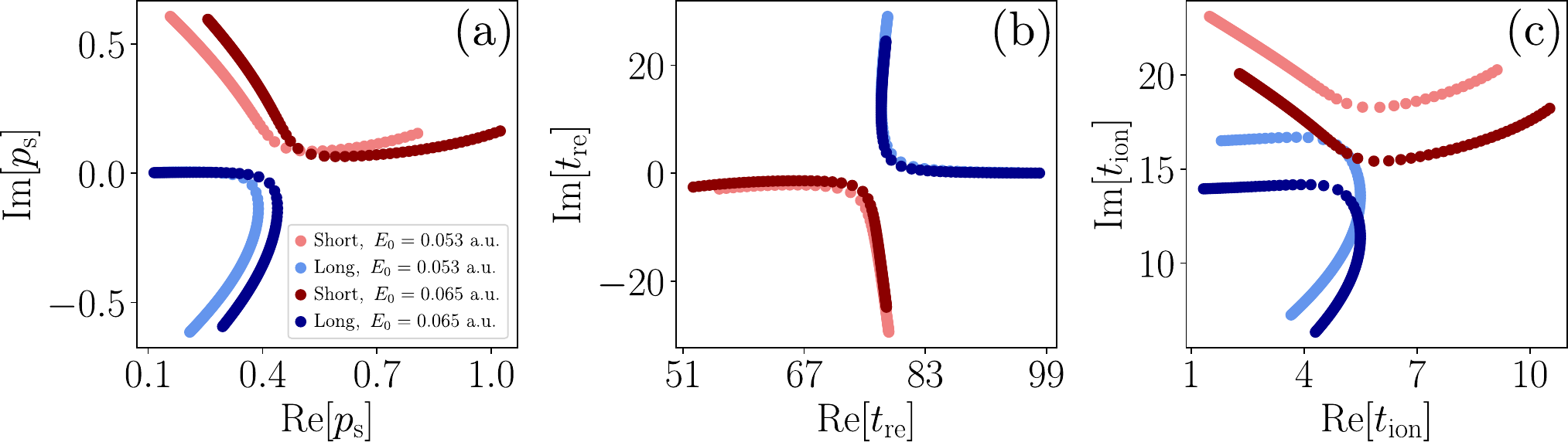}
	\caption{Real and imaginary components of the canonical momentum (panel (a)), recombination time (panel (b)), and ionization time (panel (c)) for both short (red curve) and long (blue curve) trajectories. The lighter curves represent results for $E_0 = 0.053$ a.u., while the darker curves for $E_0 = 0.065$ a.u., with $\omega_L = 0.057$ a.u.~in both cases.}
	\label{Fig:App:Trajectories}
\end{figure}

In Fig.~\ref{Fig:App:Trajectories} we show some electronic trajectories computed as specified before. The short trajectories are represented in red, while the long ones in blue.~With the lighter colors, we display the case where $E_0 = 0.053$ a.u., while the darker colors display the $E_0 = 0.065$ a.u.~case.~In both cases, we set the frequency to $\omega_L = 0.057$ a.u., though similar results are obtained with varying frequencies. Each of the points in this plot represent a different harmonic order, which extend from the 15th harmonic to the 70th for both cases. Obtaining solutions below the 15th harmonic is not always guaranteed, as the saddle-point approximation might not be adequate, though this is something depending on the field strength under consideration as well as the frequency.~However, to ensure that for all the values under consideration---$E_0 \in [0.04, 0.065]$ a.u.~and $\omega_L \in [0.04,0.057]$ a.u.---we were able to find two solutions corresponding to the long and short trajectories, we set the harmonic lower bound to the 15th harmonic order.

\begin{figure}
	\centering
	\includegraphics[width=0.8\textwidth]{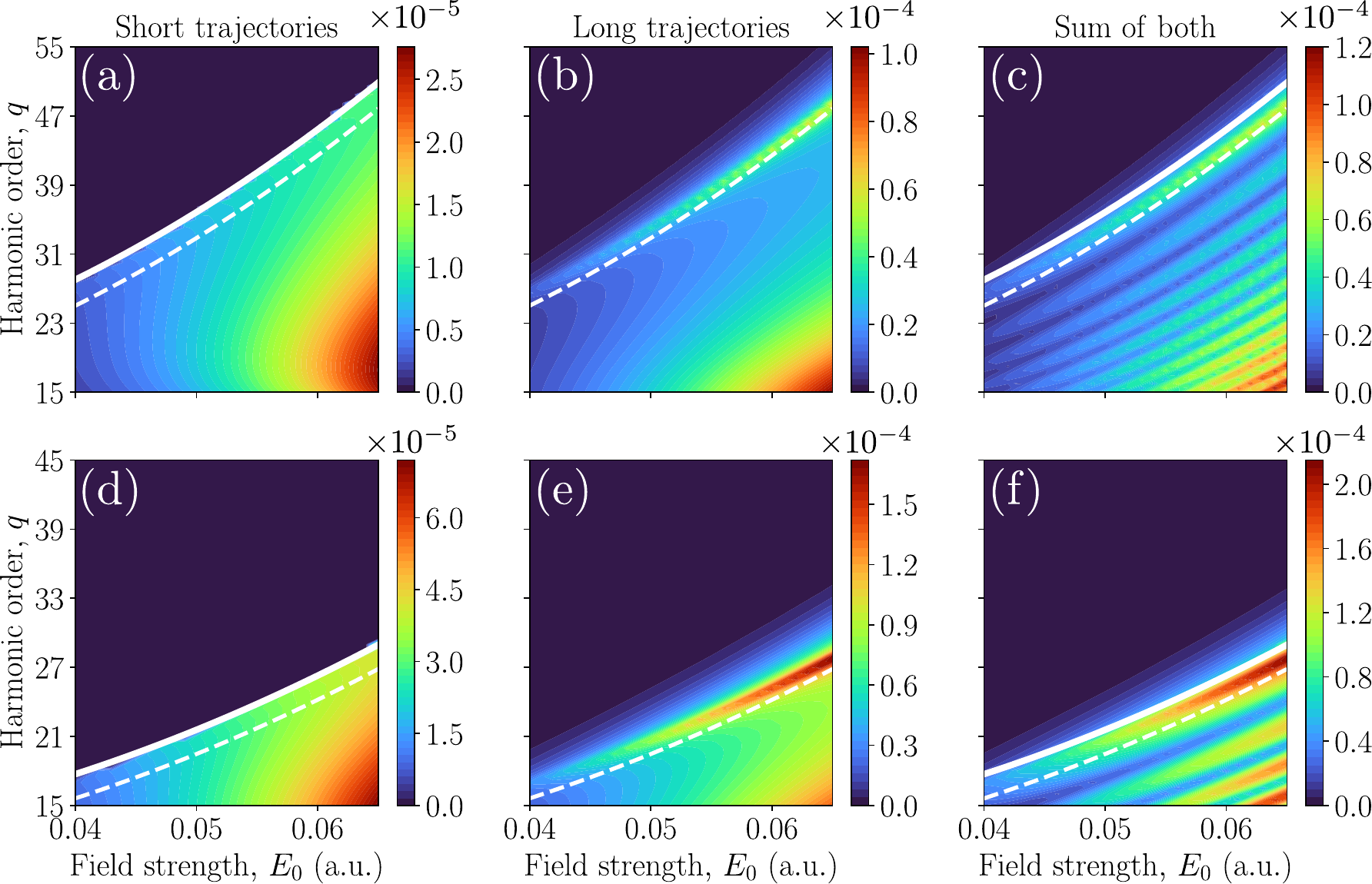}
	\caption{Values of $|M(\boldsymbol{\theta}_s)|^2$ computed using the dipole matrix elements as specified in Eq.~\eqref{Eq:App:mel:r}. Panel (a) shows the results for short trajectories, panel (b) for long trajectories, and panel (c) for their joint combination. The upper row corresponds to calculations with $\omega_L = 0.045$ a.u., while the lower row uses $\omega_L = 0.057$ a.u.~instead.}
	\label{Fig:App:HHG:Amps}
\end{figure}

While the saddle-point approximation is one of the most widely used methods for analyzing strong-field physics phenomena, it can sometimes fall short of capturing all the nuances involved. For instance, in our analysis, we have not accounted for the influence of the Coulomb potential exerted by the nucleus on the ionized electron, which could affect the electron's motion.~This seems like a reasonable approximation since the Coulomb force decreases quadratically with distance, and the strong applied field quickly drives the electron away from the parent ion.~However, problems can arise when using expressions obtained from derivations using Coulomb potentials within $M(p_s,t_{\text{re}},t_{\text{ion}})$. For example, when substituting the saddle-points computed through Eqs.~\eqref{Eq:App:Ionization} to \eqref{Eq:App:Recombination} into $M(p_s,t_{\text{re}},t_{\text{ion}}$ and using Coulomb-based expressions for the transition matrix elements between the ground and continuum states of the position operator $\hat{r}$, divergences may occur. 

To circumvent these issues, we instead use Gaussian-bound states to represent the electronic ground state and express the transition matrix elements as~\cite{lewenstein_theory_1994,nayak_saddle_2019}
\begin{equation}\label{Eq:App:mel:r}
	\mel{\text{g}}{\hat{r}}{v}
		= i 
			\bigg(\dfrac{1}{\pi \alpha}\bigg)^{3/4}
			\dfrac{v}{\alpha}
			\exp[-\dfrac{v^2}{2\alpha}],
\end{equation}
where $\alpha = 0.8 I_p$. 

Using this expression, we show in Fig.~\ref{Fig:App:HHG:Amps} the absolute value of the unnormalized probability amplitude $M(p_s,t_{\text{re}},t_{\text{ion}}$ for varying field strengths and different harmonic orders. More specifically, in the top row (panels (a) to (c)) we use $\omega_L = 0.045$ a.u., while for the bottom row (panels (d) to (f)) we use $\omega_L = 0.057$ a.u. instead. The dashed white curve represents the position of the HHG cutoff while the solid curve shows the position of the Stokes transition, occurring when $\text{Re}[S^{(q)}_{\text{sc}}]$ becomes equal for both short and long trajectories~\cite{figueira_de_morisson_faria_high-order_2002,milosevic_role_2002,pisanty_imaginary_2020}.~At this point, we neglect the contribution of short trajectories because, typically later, at the anti-Stokes transition---where $\text{Im}[S^{(q)}_{\text{sc}}]$ becomes equal for both short and long trajectories---the contribution of short trajectories start to diverge. While abruptly neglecting the contribution of short trajectories can lead to discontinuities in the value of $M(p_s,t_{\text{re}},t_{\text{ion}})$ as a function of $\omega_q$, more sophisticated methods can provide smoother transitions~\cite{figueira_de_morisson_faria_high-order_2002,milosevic_role_2002,pisanty_imaginary_2020}.~However, in our case, we restrict ourselves to the simplest scenario, as this is not the main focus of the work.

\section{About the limit of vanishing displacements due to the electron trajectories}\label{App:Product:coh}
Starting from Eq.~\eqref{Eq:ground:state:cond}, and considering the limit where $\boldsymbol{\delta}(\boldsymbol{\theta}_s) \to 0$ for both short and long trajectories, we arrive at
\begin{equation}\label{Eq:App:ground:state}
	\lvert \bar{\Phi}_{\text{g}}(t) \rangle
		= 
		\Bigg[
				\mathbbm{1}
				+ \sum_{q}^{q_c}\sum_{\{\boldsymbol{\theta_s}\}}
				\bar{M}_q(\boldsymbol{\theta}_s) 
				\hat{a}^\dagger_q
		\Bigg] \ket{\bar{0}},
\end{equation}
which, having in mind that $\hat{a}_q\ket{0} = 0$, can be also written as
\begin{equation}
	\lvert \bar{\Phi}_{\text{g}}(t) \rangle
	= 
	\Bigg[
		\mathbbm{1}
			+ \sum_{q}^{q_c}\sum_{\{\boldsymbol{\theta_s}\}}
				\Big(
					\bar{M}_q(\boldsymbol{\theta}_s) 
					\hat{a}^\dagger_q
					- \bar{M}^*_q(\boldsymbol{\theta}_s) 
					\hat{a}_q
				\Big)
		\Bigg] \ket{\bar{0}}.
\end{equation}

Given that $\bar{M}_q(\boldsymbol{\theta}_s)$ is very small irrespective for all considered harmonic modes, due to HHG being a perturbative process (in the context of perturbation theory), the state above can be approximated by
\begin{equation}
		\lvert \bar{\Phi}_{\text{g}}(t) \rangle
			\approx \exp[\sum_{q}^{q_c}\sum_{\{\boldsymbol{\theta_s}\}}
				\Big(
					\bar{M}_q(\boldsymbol{\theta}_s) 
					\hat{a}^\dagger_q
					- \bar{M}^*_q(\boldsymbol{\theta}_s) 
					\hat{a}_q
				\Big)]\ket{\bar{0}}
			= \prod_{q}^{q_c} 
				\exp[\sum_{\{\boldsymbol{\theta_s}\}}
					\Big(
						\bar{M}_q(\boldsymbol{\theta}_s) 
						\hat{a}^\dagger_q
						- \bar{M}^*_q(\boldsymbol{\theta}_s) 
						\hat{a}_q
					\Big)]
				\ket{\bar{0}},
\end{equation}
which corresponds to the multimode displacement operator $\hat{\vb{D}}(\boldsymbol{\chi}) = \prod_{q}^{q_c}\hat{D}_q(\chi_q)$, with $\chi_q = \sum_{\boldsymbol{\theta}_s}\bar{M}_q(\boldsymbol{\theta}_s)$.~Since this operator acts locally on each harmonic mode independently, we can conclude that the entanglement features in Eq.~\eqref{Eq:App:ground:state} vanish in the limit of small displacements. It is worth noting that these results are in agreement with those of Refs.~\cite{lewenstein_generation_2021,rivera-dean_strong_2022,stammer_quantum_2023}, obtained when neglecting the back-action of the electron on the field modes.~Similar to these references, the multiatom approach considered in this work would result in $\hat{\vb{D}}(\boldsymbol{\chi}) = \hat{\vb{D}}(N_{\text{at}}\boldsymbol{\chi})$, where the displacements are scaled by an additional prefactor.~However, this scaling does not affect the entanglement features between the field modes.

\section{About the numerical evaluation of the quantum optical observables and entanglement measures}\label{App:Num:Eval}
The numerical evaluation in this work can be divided into two parts: the saddle-point analysis and the non-classical analysis. As mentioned in Appendix~\ref{App:Saddle:Point}, the former was conducted in \texttt{Mathematica} by using the \texttt{EPToolbox} package~\cite{pisanty_episantyeptoolbox_2018}, while the latter was performed in \texttt{Python}, primarily utilizing the \texttt{QuTiP} package~\cite{johansson_qutip_2012,johansson_qutip_2013}. The numerical integration of the displacement $\delta_{q}(\boldsymbol{\theta}_s)$ in Eq.~\eqref{Eq:App:displacement}, which under the saddle-point approximation involves complex-valued integration limits, was carried out using the \texttt{quad} function of the \texttt{mpmath}~\cite{mpmath} package. To ensure the accuracy of this function, we compared the results with those obtained from \texttt{Mathematica}, where the indefinite integral was solved exactly, and numerical values were introduced afterwards for comparison.

The quantum optical state in Eq.~\eqref{Eq:two:harm:state} was implemented in the Fock basis, as per standard in \texttt{QuTiP}, by first constructing the operator kernel for $N_{\text{at}} = 1$.~To account for the mullti-atom effect described in the main text, this kernel was multiplied by itself $N_{\text{at}}-1$ times.~While the Fock basis is inherently infinite-dimensional, a cutoff must be introduced to make numerical calculations feasible.~In this work, we set the cutoff to $n_{q=1,\text{cutoff}} = 180$ for the fundamental mode and $n_{q=1,\text{cutoff}} = 20$ for the harmonic modes.~Convergence was verified by considering slightly higher cutoffs for the highest field strength considered in this study. 

With the numerical expression of the quantum optical state, the Wigner function was evaluated using built-in \texttt{QuTiP} functions, specifically the \texttt{wigner} function. While the linear entropy can be straightforwardly computed once the full quantum optical state is available, evaluating the logarithmic negativity requires a slightly more involved procedure. In this case, we computed the partial transpose with respect to one of the subsystems using the \texttt{partial\_transpose} function in \texttt{QuTiP}.~Then, we evaluated the eigenvalues of the resulting matrix form of the state using the built-in functions of the \texttt{Numpy} package~\cite{harris2020array}, and retained only the negative ones. To verify the accuracy of this method, we benchmarked the results obtained through them against those resulting from Bell states.

\end{document}